\documentclass[pra, longbibliography, aps, nofootinbib]{revtex4-2}   	

\usepackage{stmaryrd} 			
\usepackage{geometry}                		
\usepackage{graphicx}
\usepackage{amssymb, amsmath}
\usepackage{subcaption}
\usepackage{dsfont}
\usepackage{hyperref}

\usepackage[htt]{hyphenat} 		

\usepackage{amsthm}   			
\newtheorem{axiom}{Axiom}

\usepackage[titletoc]{appendix} 	  

\makeatletter
\newcommand{\specialcell}[1]{\ifmeasuring@#1\else\omit$\displaystyle#1$\ignorespaces\fi}
\makeatother

\makeatletter
\newcommand*\bigcdot{\mathpalette\bigcdot@{.5}}
\newcommand*\bigcdot@[2]{\mathbin{\vcenter{\hbox{\scalebox{#2}{$\m@th#1\bullet$}}}}}
\makeatother

\newcommand{\dersub}[2]{\overset{\,\,\scriptscriptstyle{(\hspace{-.8pt}\raisebox{0.8pt}{\text{\large{.}}} \hspace{-.8pt})^{#2}}}{#1}}

\newcommand{\I}{\text{I}}
\newcommand{\II}{\text{II}}

\renewcommand{\d}{\text{d}}

\newcommand{\lab}{\text{lab}}
\newcommand{\rot}{\text{rot}}
\newcommand{\eff}{\text{eff}}

\newcommand{\var}{\text{var}}
\newcommand{\eig}{\text{eig}}
\newcommand{\exact}{\text{exact}}
\newcommand{\ex}{\text{ex}}

\newcommand{\n}{\textbf{n}}
\newcommand{\gate}{\text{gate}}

\newcommand{\RWA}{\text{RWA}}
\renewcommand{\O}{\mathcal{O}}
\renewcommand{\H}{\mathcal{H}}

\begin{document}

\title{Refuting a Proposed Axiom for Defining the Exact Rotating Wave Approximation}
\author{Daniel Zeuch$^1$ and David P.~DiVincenzo$^{1,2}$}
\affiliation{$^1$Peter Gr\" unberg Institut, Theoretical Nanoelectronics, Forschungszentrum  J\" ulich, D-52425 J\"ulich, Germany \\
$^2$Institute for Quantum Information, RWTH Aachen University, 52062 Aachen, Germany}
\date{\today}

\begin{abstract}
For a linearly driven quantum two-level system, or qubit, sets of stroboscropic points along the cycloidal-like trajectory in the rotating frame can be approximated using the exact rotating wave approximation introduced in arXiv:1807.02858.  That work introduces an effective Hamiltonian series $\H_{\eff}$ generating smoothed qubit trajectories; this series has been obtained using a combination of a Magnus expansion and a Taylor series, a Magnus-Taylor expansion.  Since, however, this Hamiltonian series is not guaranteed to converge for arbitrary pulse shapes, the same work hypothesizes an axiomatic definition of the effective Hamiltonian.  The first two of the proposed axioms define $\H_{\eff}$ to (i) be analytic and (ii) generate a stroboscopic time evolution.  In this work we probe a third axiom---motivated by the smoothed trajectories mentioned above---namely, (iii) a variational principle stating that the integral of the Hamiltonian's positive eigenvalue taken over the full pulse duration is minimized by this $\H_{\eff}$.  We numerically refute the validity of this third axiom via a variational minimization of the said integral.  
\end{abstract}


\maketitle

\tableofcontents

\newpage

\section{Introduction}
\label{introduction}

Consider a quantum two-level system, or qubit, which is coupled to a linearly-polarized drive treated classically.  This problem, which has been considered by Bloch and Siegert \cite{bloch40}, is of current interest due to its applicability to the field of quantum information processing \cite{nielsen10}.  In this setting, single-qubit gates need to be carried out with high precision by shaped pulses.  Optimal pulse shapes, which correspond to specific envelope functions, are of often required to satisfy multiple constraints \cite{motzoi09} and therefore need to be synthesized by way of numerical search.  Such a search can be streamlined by the ability to predict the driven qubit's time evolution using a high-precision approximation that is easy to integrate numerically.  

Given a resonance frequency of the qubit, $\omega_0$, and a drive frequency, $\omega$, the Hamiltonian of the driven-qubit system reads\footnote{While adding a term proportional to the identity to this Hamiltonian does not change the dynamics of the system, we consider a traceless Hamiltonian for simplicity in our analysis. \label{foot:traceless}} ($\hbar = 1$)
\begin{eqnarray}
	\mathcal H_{\lab}(t) &=& \frac{\omega_0}2 \sigma_z + \frac{H_1(t)}{2} \cos(\omega t + \phi) \sigma_x 
	\label{Hlab0}\\
			&=& \frac{\omega}2 \sigma_z + \frac{H_1(t)}{2} \cos(\omega t) \sigma_x, \qquad \qquad (\omega_0 = \omega, \phi = 0).
	\label{Hlab}
\end{eqnarray}
Here $H_1(t)$ is the time-dependent drive amplitude, and $\sigma_x$, $\sigma_y$, $\sigma_z$ are the Pauli matrices.  For the Hamiltonian (\ref{Hlab0}) we assume a constant phase offset, $\phi$, and small detuning, $\Delta = \omega_0 - \omega$, with $\Delta \ll \omega$.  In the present study we often consider the Hamiltonian (\ref{Hlab}), which corresponds to the special case of resonant driving, $\Delta \equiv \omega_0 - \omega = 0$, and zero phase offset $\phi=0$.  

The relative magnitudes of the amplitude $H_1(t)$ and the qubit frequency $\omega$, the two central parameters in the above Hamiltonian, can be used to define different parameter regimes.  Here, we focus on the regime of relatively weak to strong driving in which $|H_1(t)| \lesssim \omega$, for which it is useful to transform the above Hamiltonians from the laboratory frame of reference to a rotating frame associated with the drive.  This latter frame rotates about the $z$ axis with the drive frequency $\omega$ and is defined by the standard transformation $\mathcal H_{\rot} = \tilde U^{\dagger} \mathcal H_{\lab} \tilde U - i \tilde U^\dagger \frac{\partial}{\partial t}\tilde U$ \cite{messiah1964quantum} with the unitary operator $\tilde U(t) = e^{-i \omega t \sigma_z/2}$,
\begin{eqnarray}
	\mathcal H_{\rot}(t) &\stackrel{(\ref{Hlab0})}{=}& \frac{H_1(t)}4 ( \cos(\phi)\sigma_x + \cos(2\omega t + \phi) \sigma_x + \sin(\phi)\sigma_y - \sin(2\omega t + \phi)\sigma_y) + \frac\Delta2 \sigma_z
	\label{Hrot0}\\
		&\stackrel{(\ref{Hlab})}{=}& \frac{H_1(t)}4 ( \sigma_x + \cos(2\omega t) \sigma_x - \sin(2\omega t)\sigma_y), \quad \qquad (\Delta=0, \ \phi=0).
	\label{Hrot}
\end{eqnarray}
While the drive in the lab frame has a period of $2\pi/\omega$, note that the drive period in the rotating frame is
\begin{equation}
	t_c = \pi/\omega.
	\label{tc}
\end{equation}

Since the rotating-frame Hamiltonian $\mathcal H_{\rot}(t)$ does not commute with itself at arbitrary times, it is a nontrivial problem to compute its time evolution analytically.  Further note that non-commuting terms in the rotating-frame Hamiltonian $\mathcal H_{\rot}(t)$ vary on the time scale of $1/\omega$.  This time scale is assumed small compared to the Rabi frequency ($\sim \max(|H_1(t)|)$), since for most realistic pulses the amplitude fulfills $|H_1(t)| \leq 0.1 \omega$.  This fast time dependence implies that $\mathcal H_{\rot}(t)$ cannot be integrated very easily.  This problem is often circumvented by using the rotating wave approximation (RWA) \cite{cohen98}.  The Hamiltonian in the RWA is obtained by taking the rotating-frame Hamiltonian and neglecting the oscillatory terms.  The usefulness of the RWA is that the Hamiltonian in this approximation varies relatively slowly in time, rendering it easy to integrate.  However, the RWA only gives relatively accurate results for very weak drives with $|H_1(t)| \ll \omega$.  

The driven-qubit problem described above has been studied using Floquet's theorem (see, e.g., Refs.~\cite{shirley65, aravind84, peskin93, drese99, mananga11, novicenko17, schmidt18}), the dressed-state formalism \cite{cohen73} and the Magnus expansion (see, e.g., work on nuclear magnetic resonance \cite{haeberlen68, evans68, waugh68}, or more recent studies \cite{casas01, blanes09, mananga11, bukov2015universal} in which Floquet theory and the Magnus expansion have been combined).  Not long ago, it has been shown that the time-dependent Schroedinger equation can be solved using the path-sum method \cite{giscard15}, and its applicability to the driven qubit problem constant drive amplitudes has been demonstrated in Ref.~\cite{giscard2019general}.  For a more complete literature review see Ref.~\cite{zeuch18}.  

A recent development in the subject of periodically-driven quantum systems is the introduction of the exact rotating wave approximation \cite{zeuch18}, which can be used to accurately predict the time evolution even for strong drives with $|H_1(t)| \lesssim \omega$.  This theory is based on a novel method for time-dependent perturbation theory called the Magnus-Taylor expansion \cite{zeuch18}.  The time evolution in the exact RWA is generated by an \textit{effective Hamiltonian}, which, when compared to the exact Hamiltonian, varies only slowly in time and can therefore be integrated with similar ease as the RWA Hamiltonian.  We note that Ref.~\cite{varvelis19} applies this effective Hamiltonian to the problem of designing quantum gates for singlet-triplet spin qubits \cite{cerfontaine14}.  

Here we are concerned with the definition of the effective Hamiltonian, denoted $\H_\eff$.  In its original derivation \cite{zeuch18}, $\H_\eff$ is formulated as a series, whose convergence, however, is not always guaranteed.  In Ref.~\cite{zeuch18} it has therefore also been surmised that the effective Hamiltonian can be alternatively defined via an axiomatic definition whose motivation is based on the qualitative features related to the stroboscopic time evolution.  In the present paper we propose, study and give numerical evidence against a particular variant of such an axiomatic definition.

\subsection{Exact Rotating Wave Approximation}
\label{exactRWA}

The problem of finding the time evolution for the driven qubit is captured by the Schroedinger equation,
\begin{eqnarray}
	- i \partial_t |\psi(t)\rangle = \H(t) |\psi(t)\rangle,
	\label{schroedinger}
\end{eqnarray}
where for our problem the exact Hamiltonian is given by one of the rotating-frame Hamiltonians (\ref{Hrot0}) or (\ref{Hrot}).  The solution to the Schroedinger equation can be formally expressed via the time evolution operator for initial and final times $t_i$ and $t$, respectively,
\begin{equation}
	U(t, t_i) = \mathcal{T}e^{-i\int_{t_i}^{t} \d \tau \mathcal H(\tau)} = e^{-i \overline{\mathcal H} (t-t_i)}.
	\label{Ugeneric}
\end{equation}
Here the unitary operator $U$ is first written in the usual form featuring the time ordering operator $\mathcal{T}$.  We also express the time evolution operator as a true exponential function using the Magnus expansion \cite{magnus1954, ernst87, waugh07}, in which the quantity $\overline {\mathcal H}$, also referred to as a Magnus Hamiltonian, can be understood as a type of Hamiltonian average on the interval $[t_i, t]$.  This average is usually given as series of integral terms of commutators of the Hamiltonian with itself at different times, and the first three terms of this series are given explicitly in Appendix \ref{magnus_appendix}.  A formal solution to the Schroedinger equation (\ref{schroedinger}) then reads
\begin{eqnarray}
	|\psi(t)\rangle = U(t, t_i)|\psi(t_i)\rangle.
	\label{solution}
\end{eqnarray}


Note that even for a constant envelope function $H_1(t) \equiv H_1$ the computation of the time evolution operator (\ref{Ugeneric}) is nontrivial because the full rotating-frame Hamiltonian $\H_\rot(t)$ does not commute with itself at different times $t$ and $t'$, $[\H_\rot(t), \H_\rot(t')] \neq 0$.  In contrast, the RWA Hamiltonian,
\begin{eqnarray}
	\H_{\RWA}(t) &\stackrel{(\ref{Hrot0})}{=}& \frac{H_1(t)}4 ( \cos(\phi)\sigma_x + \sin(\phi)\sigma_y) + \frac\Delta2 \sigma_z \\ 
			&\stackrel{(\ref{Hrot})}{=}& \frac{H_1(t)}4 \sigma_x,  \quad \qquad \qquad\qquad\qquad (\Delta=0, \ \phi=0).
	\label{HRWA}
\end{eqnarray}
does commute with itself at different times for either case of zero detuning $\Delta=0$ or a constant field amplitude $H_1(t)$.  [As noted above, the RWA Hamiltonian is obtained by neglecting the oscillating terms in the rotating frame Hamiltonian given above.]  In either case the computation of the time evolution operator (\ref{Ugeneric}) for this approximation simplifies greatly since the time ordering operator $\mathcal T$ can be neglected.  For the simplest case of a constant amplitude the RWA Hamiltonian itself is a constant, and we have
\begin{eqnarray}
	\qquad \qquad \qquad U_{\RWA}(t, t_i) = e^{-i\int_{t_i}^{t} \d \tau  \H_{\RWA}} = e^{-i \H_{\RWA} (t-t_i)},	\qquad \qquad (H_1(t) \equiv H_1).
	\label{URWA}
\end{eqnarray}

As noted above, when the ratio $|H_1(t)|/\omega$ is appreciable, the usage of the RWA is not justfied for many applications requiring high-precision predictions of the qubit's time evolution.  The perhaps most famous correction to the RWA is the Bloch-Siegert shift \cite{bloch40}, which is proportional to $H_1(t)^2/\omega$ at lowest order in $1/\omega$.  A Hamiltonian beyond the RWA may then be written as follows,
\begin{eqnarray}
	\H_{\RWA, \text{improved}}(t) = \frac{H_1(t)}4 \sigma_x - \frac{H_1(t)^2} {32\omega} \sigma_z.
	\label{BS}
\end{eqnarray}
As has been pointed out recently \cite{zeuch18}, while this Hamiltonian (\ref{BS}) is a systematic improvement for constant drive envelopes, this is not the case for arbitrary envelopes.  This is because $\H_{\RWA, \text{improved}}$ does not capture a correction term proportional to $\dot H_1/\omega$, which is of importance since it is on the same order in $1/\omega$ as the Bloch-Siegert shift.  In the exact RWA, this term is part of an \emph{effective Hamiltonian} that has been derived in Ref.~\cite{zeuch18} using the Magnus-Taylor expansion mentioned above.  Reference \cite{zeuch18} introduces an effective Hamiltonian as a series expansion in $1/\omega$,
\begin{eqnarray}
	\H_{\eff}(t; \beta_0) = \sum_{k=0}^{\infty} h_{k}(t; \beta_0) (1/\omega)^k.
	\label{HeffSeries}
\end{eqnarray}
The operator function $h_{k=0}$ corresponds to the case of the RWA, i.e., $h_{k=0}(t; \beta_0)=\H_{\RWA}(t)$.  For a given $k>0$, the operator $h_{k}$ can be computed using the recursion relation given by Eq.~(59) in Ref.~\cite{zeuch18}.  The first five terms of the series (\ref{HeffSeries}) for the special-case rotating-frame Hamiltonian (\ref{Hrot}) are given explicitly in Appendix \ref{effH_appendix}.  

This Hamiltonian series constitutes a set of correction terms to the usual RWA Hamiltonian given above, and it can be obtained up to arbitrary order in $1/\omega$.  Assuming this series converges, the effective Hamiltonian results in a stroboscopic time evolution, that is, it generates effective qubit trajectories that agree with the exact trajectory at periodic points in time.  Note that this effective Hamiltonian depends not only on time $t$ but also on a gauge parameter, $\beta_0$, whose role is explained further below.  

To give an example, the effective Hamiltonian for the system described by the rotating frame Hamiltonian (\ref{Hrot}) is
\begin{eqnarray}
		\H_\eff(t; \beta_0) &=& \frac{H_1}{4}\sigma _x + \frac{H_1^2}{32 \omega}(1-2 \cos\beta_0)\sigma_z + \frac{\dot H_1}{8 \omega}(\sin\beta_0\sigma_x+\cos\beta_0\sigma_y) + \mathcal O(1/\omega^2) \label{Heff_main} \\
			&\stackrel{(\beta_0=0)}{=}&\frac{H_1}{4}\sigma _x - \frac{H_1^2}{32 \omega}\sigma_z + \frac{\dot H_1}{8 \omega}\sigma_y + \mathcal O(1/\omega^2),
\end{eqnarray}
which is here given only up to first order in $1/\omega$.  This Hamiltonian gives a systematic prediction of the time evolution of the driven qubit for time-dependent drive envelopes $H_1(t)$.  Note that for constant $H_1(t) = H_1$ this effective Hamiltonian for $\beta_0=0$ reduces to the improved Hamiltonian (\ref{BS}), which includes the Bloch-Siegert shift.  

\begin{figure}
	\includegraphics[width = \columnwidth]{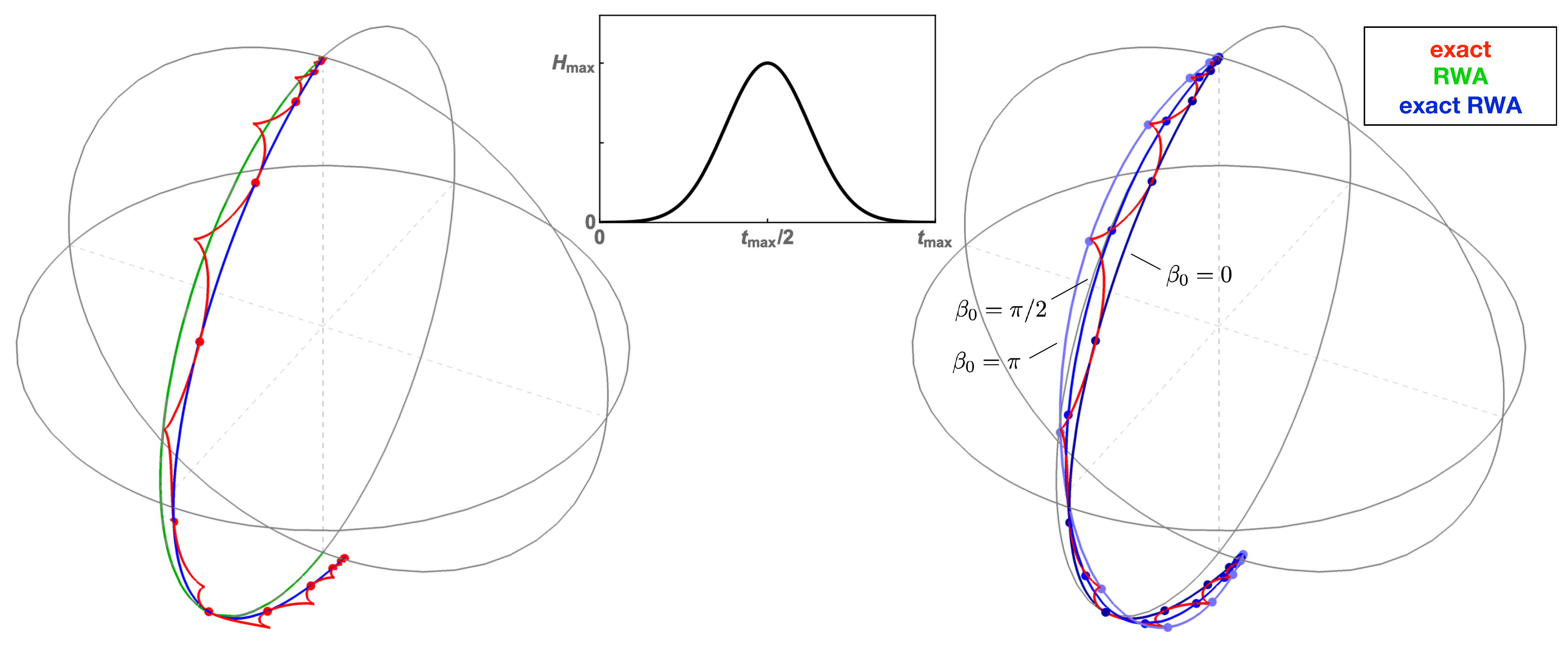}
	\caption{Various qubit trajectories in the rotating frame compared and contrasted to one another.  The initial state of the qubit is $|\psi(t=0)\rangle = |0\rangle$.  The amplitude and width of the Gaussian envelope function (give function explicitly) are chosen such that the RWA trajectory, shown in green on the left-hand side (LHS), (i) corresponds to a $\pi$-pulse and (ii) is visibly inaccurate.  As opposed to the simple RWA path, the exact trajectory (red) describes a cycloidal-like path.  The LHS also shows the trajectory (blue) corresponding to the effective Hamiltonian of the exact RWA.  On the right-hand side we show how three different \emph{exact RWA} trajectories for $\beta_0 = 0$, $\pi/2$ and $\pi$ (as indicated in the figure) match with the exact path at stroboscopic points in time.}
	\label{trajectories}
\end{figure}

Figure \ref{trajectories} illustrates the general behavior of the exact RWA by means of various qubit Bloch-sphere trajectories for a $\pi$-pulse in the RWA with a Gaussian envelope function $H_1(t)$, which fulfills $\int_{0}^{t_\text{gate}} \text{d}\tau\, H_1(\tau) = 2\pi$ with the pulse duration $t_\text{gate}$.  For this choice, the time evolution operator in the RWA, given in Eq.~(\ref{URWA}), results in a \textsc{not} gate, $U_{\RWA}(t_\gate) \propto \sigma_x$.  Shown are various solutions $|\psi(t)\rangle$ with $t\in[0, t_\gate]$ to the Schroedinger equation (\ref{schroedinger}) with initial condition $|\psi(t=0)\rangle = |0\rangle$, that is, at initial time $t_i = 0$ the qubit is initialized to the north pole of the Bloch sphere.  Each shown trajectory corresponds to a certain Hamiltonian as detailed below.  

The exact qubit trajectory $|\psi_{\exact}(t) \rangle = U_{\exact}(t, 0)|0\rangle$ [cf.~the solution (\ref{solution}) to the Schroedinger equation], shown in red, is generated by the exact Hamiltonian (\ref{Hrot}).  Its time evolution operator of the generic form (\ref{Ugeneric}) is given by
\begin{equation}
	U_{\exact}(t, t_i) = \mathcal{T}e^{-i\int_{t_i}^{t} \d \tau \mathcal H_\rot(\tau)}.
	\label{Uexact}
\end{equation}
The corresponding trajectory follows cycloidal-like motions known as Bloch-Siegert oscillations, which are due to the terms in the exact Hamiltonian (\ref{Hrot}) that oscillate at twice the drive frequency $\omega$.  In contrast, the RWA trajectory, generated only by the operator $\sigma_x$ [see the RWA Hamiltonian (\ref{HRWA})] is a simple $x$-axis rotation.  As becomes clear from the trajectories shown on the left-hand side (LHS) of Fig.~\ref{trajectories}, this RWA trajectory, shown in green in the figure, makes significant errors in predicting the exact trajectory for the chosen, relatively hard drive ($H_1(t)/\omega \lesssim 0.1$).  The LHS of Fig.~\ref{trajectories} also shows a blue qubit trajectory due to the effective Hamiltonian.  As described above, the effective and exact trajectories agree at stroboscopic points indicated with bullets in the figure.  The time difference between these points is the period of the drive (\ref{tc}), or $t_c = \pi/\omega$, which is equal to the period of the Bloch-Siegert oscillations.

As noted above, the effective Hamiltonian $\H_\eff=\H_\eff(t; \beta_0)$ depends on a gauge parameter denoted $\beta_0$.  This gauge parameter enables one to choose different sets of stroboscopic points at which the effective and exact trajectories agree; these sets are given by
\begin{eqnarray}
	\{t_0, t_0 \pm t_c, t_0 \pm 2t_c, \ldots\}.
	\label{sets}
\end{eqnarray}
Here the constant time offset $t_0$ is chosen $t_0 \in [0, t_c)$, where $t_c$ given in Eq.(\ref{tc}) denotes the period of the drive in the rotating frame.  We denote the intervals
\begin{eqnarray}
	[t_0 + n t_c, t_0 + (n+1) t_c)
	\label{MagnusInterval}
\end{eqnarray}
as \emph{Magnus intervals}.  The offset $t_0$ is then related to the gauge parameter $\beta_0$ via the drive period $t_c = \pi/\omega$, that is,
\begin{eqnarray}
	\beta_0 = 2\pi t_0/t_c = 2\omega t_0, \qquad \qquad \beta_0 \in [0, 2\pi).
	\label{beta0}
\end{eqnarray}
Reference \cite{zeuch18} refers in this context to a gauge degree of freedom because both the starting and endpoints for a drive pulse are left unchanged\footnote{For this statement to be exact, one needs to implement so-called kick operators \cite{zeuch18}.} when varying $\beta_0$.  This behavior is exemplified on the right-hand side of Fig.~\ref{trajectories}, where three different effective qubit trajectories for $\beta_0 = 0$, $\pi/2$ and $\pi$ are shown in different shades of blue.  In this plot points of agreement are again indicated by bullets.  

With reference to the generic time evolution described by Eq.~(\ref{Ugeneric}), the effective evolution operator from an initial time $t_i = t_0 + m t_c$ for some integer $m$ is written in an extended notation,
\begin{equation}
	U_{\beta_0}(t, t_i = t_0 + m t_c) = \mathcal{T} e^{-i\int_{t_0 + m t_c}^{t} \d \tau \mathcal H_{\eff}(\tau; \beta_0)},
	\label{Ueff}
\end{equation}
in which the dependence on the gauge parameter $\beta_0$ is given explicitly.  Note that the choice of this gauge parameter $\beta_0$ determines the initial time $t_i$ up to an integer multiple of the drive period $t_c$, since the time $t_0$ is related to $\beta_0$ via Eq.~(\ref{beta0}).  The stroboscopic time evolution means that this time evolution operator $U_{\beta_0}(t, t_i)$ agrees with the exact time evolution, which is described by Eq.~(\ref{Ugeneric}) with $\H = \H_\rot$ and the same initial time $t_i$, at the final times (\ref{sets}).  

It is a premise of the derivation in Ref.~\cite{zeuch18} that the effective Hamiltonian depends on time solely through the envelope function $H_1(t)$ and its derivatives, or 
\begin{eqnarray}
	\H_\eff(t) = \H_\eff(H_1(t), \dot H_1(t), \ddot H_1(t), \ldots).
	\label{HEff_dependence}
\end{eqnarray}
For the special case of constant $H_1(t) \equiv H_1$ it then follows that the effective Hamiltonian is time-independent, $\H_\eff(t; \beta_0) = \H_\eff(\beta_0)$.  Since time ordering can then be ignored, the effective time evolution operator (\ref{Ueff}) greatly simplifies to
\begin{eqnarray}
	\qquad \qquad \qquad U_{\beta_0}(t, t_i) = e^{-i\int_{t_i}^{t} \d \tau  \H_{\eff}(\beta_0)} = e^{-i \H_{\eff}(\beta_0) (t-t_i)},	\qquad \qquad (H_1(t) \equiv H_1).
	\label{Ueff_const}
\end{eqnarray}
For constant amplitudes the time evolution in the exact RWA is thus computed as easily as in the regular RWA, cf.~Eq.~(\ref{URWA}).



Reference \cite{zeuch18} presents a derivation of the effective Hamiltonian series.  This is done by employing a new method for time-dependent perturbation theory that combines the Magnus expansion \cite{magnus1954, ernst87, waugh07} of the time evolution operator with a Taylor expansion of the envelope $H_1(t)$---this method has been called the Magnus-Taylor expansion \cite{zeuch18}.  As a consequence, the effective Hamiltonian is a function of not only $H_1(t)$ but also all its temporal derivatives, as also highlighted in Eq.~(\ref{HEff_dependence}).  A direct consequence of this fact is that a crucial condition for an analytic effective Hamiltonian is an analytic envelope function $H_1(t)$.  For non-analytic $H_1(t)$ the theory of the exact RWA requires the application of kick operators \cite{zeuch18}.  Here we will not need to employ such kick operators, because ignoring them only leads to minor numerical corrections that are smaller than the error due to other numerical imperfections.  

Related to the nontrivial subject of the convergence of the Magnus expansion, a caveat of the effective Hamiltonian of Ref.~\cite{zeuch18} is that the circumstances under which the approximation series for the effective Hamiltonian converges have not yet been established.  In Ref.~\cite{zeuch18} it was, however, hypothesized that there may be an alternate, \emph{axiomatic} definition of the effective Hamiltonian, which may hold even if the by the proposed calculation method does not converge.  Two axioms in that definition would be that 1) the effective Hamiltonian is an analytic function of time, and 2) its propagator agrees with the exact propagator at periodic points in time.  Since these two axioms are fulfilled not only by the effective Hamiltonian but also by the exact Hamiltonian, there needs to be at least a third axiom that sets these two Hamiltonians apart.  

The present manuscript is a compilation of research notes composed while probing such a third axiom.  This axiom can be motivated by the exact and effective trajectories shown in Fig.~\ref{trajectories}, which illustrates the notion that individual effective trajectories are significantly shorter and smoother than the exact trajectory.  This observation is the basis for the proposal of a third axiom, which suggests that the positive eigenvalue of the effective Hamiltonian is, when averaged over the entire qubit path, smaller than that of the exact Hamiltonian --- and thus perhaps also smaller than that of any other Hamiltonian generating a stroboscopic time evolution.  

Referring to Fig.~\ref{trajectories}, an effective qubit trajectory is generally visibly \emph{shorter} than its exact counterpart.  Since the length of a trajectory is related to the eigenvalue of a Hamiltonian, one proposal discussed in this manuscript is that the integral of the positive eigenvalue of the Hamiltonian over the entire pulse duration will be smaller than the same integral for any other Hamiltonian satisfying Axioms 1 and 2.  

Furthermore, Fig.~\ref{trajectories} suggests that the effective trajectories are in general significantly \emph{smoother} than its exact counterpart.  Since the smoothness is related to the change of the Hamiltonian as a function of time, our second proposal is that the integral of the positive eigenvalue of the time derivative of the Hamiltonian (again taken over the entire pulse duration) will be smaller than the same integral for any other Hamiltonian satisfying axioms 1 and 2.  

It is the purpose of this study to propose and investigate such a third axiom that may determine the effective Hamiltonian.  Our investigation is partially analytic, though the main part consists of the implementation of a numerical minimization.  

The goal of this study has been to either prove the axiomatic definition analytically, or find supporting evidence for or against it using a numerical investigation.  Since for most integrals that appear in this study an analytic solution has escaped our notice, we have tried to minimize the respective functionals numerically.  Indeed, we find that the integrals that we proposed to be minimized by the effective Hamiltonian are actually minimized by a different Hamiltonian, which also results in a stroboscopic time evolution.  This numerically-obtained finding refutes the third axiom.

\subsection{Structure of the Remainder of This Manuscript}
\label{structure}

These research notes are structured as follows.  In Sec.~\ref{prelims} we first present the proposed \emph{axiomatic} definition of the effective Hamiltonian, which is the subject of this study.  We then explain the basic method for scrutinizing this axiomatic definition.  In Sec.~\ref{minimizeEigenvalue} we compute algebraic formulas for the main integrand considered in this work, both for constant driving envelopes as well as for time-dependent envelopes.  In Sec. \ref{Results_of_Integrals} we present our numerical analysis, and we conclude in Sec.~\ref{conclusions}.  

A set of Python scripts and Mathematica notebooks used for the variational minimization described below can be found and accessed on a GitHub repository, see \href{https://github.com/zeuch/exactRWA.git}{this link}.  A set of Mathematica files that can be used for computing the effective Hamiltonians can be found in the same repository.


\section{Preliminaries}
\label{prelims}

Starting with the proposed axiomatic definition described in Sec.~\ref{axioms}, we go on to propose two particular integrands for the functional of the third proposed axiom in Sec.~\ref{integrands}.  We then introduce the details of our variational minmization in Sec.~\ref{minimization}, and mention the role of kick operators in \ref{kicks}.  In Sec.~\ref{coding} we give a short overview over the github repository containing the Mathematic notebooks and Python scripts used for this work.


\subsection{Axiomatic Definition of the Effective Hamiltonian}
\label{axioms}

This Hamiltonian generates the time evolution in the exact rotating wave approximation as described above in Sec.~(\ref{exactRWA}).  

As noted in the Introduction, the convergence of the Hamiltonian series (\ref{HeffSeries}) is not always guaranteed to converge.  

Because of this convergence problem, we propose an axiomatic definition of the effective Hamiltonian.  For this we assume that the exact Hamiltonian, given in Eq.~(\ref{Hrot}) for a linearly driven qubit in the rotating frame, itself is an analytic function in time.  Note that for the Hamiltonian (\ref{Hrot}) this is the case if the envelope function $H_1(t)$ is analytic.  It follows from the derivation of the effective Hamiltonian \cite{zeuch18} that the $\H_\eff$ is analytic in time.  This is because $\H_\eff(t; \beta_0)$ depends only on constants and on $H_1(t)$ and its derivatives (see the effective Hamiltonian terms given in Appendix \ref{effH_appendix} for example terms).  Furthermore, the effective and exact qubit trajectories coincide at equally-spaced points in time.  This is the basis for the first two axioms,
\begin{axiom}
	The effective Hamiltonian $\H_{\eff}(t; \beta_0)$ is analytic in both of its arguments, that is, in time $t$ and the gauge parameter $\beta_0$.  
	\label{axiom1}
\end{axiom}
\begin{axiom}
	The time evolution due to the effective Hamiltonian $\H_{\eff}(t; \beta_0)$ agrees with the exact time evolution at times (\ref{sets}), i.e. $t_0$, $t_0 + t_c$, $t_0 + 2t_c$, \ldots with $t_0 = \beta_0/2\omega$.  That is, the time evolution operators for the exact time evolution, Eq.~(\ref{Ugeneric}), and for the effective time evolution, Eq.~(\ref{Ueff}), must be equal at these times (\ref{sets}),
	\begin{equation}
		U_{\beta_0}(t_0 + n t_c, t_i) = \mathcal{T}e^{-i\int_{t_i}^{t_0 + n t_c} \d \tau \H_{\eff}(\tau; \beta_0)}
				\stackrel{!}{=} \mathcal{T}e^{-i\int_{t_i}^{t_0 + n t_c} \d \tau \H_{\rot}(\tau)} 	\qquad \forall n\in\mathbb{N}.
		\label{goal0}
	\end{equation}
	Here we have $t_i = t_0 + m t_c$ for an integer $m$ [recall that $\beta_0 = 2\pi t_0/t_c$ as per Eq.~(\ref{beta0})].
	\label{axiom2}
\end{axiom}

Note that Axioms \ref{axiom1} and \ref{axiom2} are not only satisfied by the effective Hamiltonian $\H_{\eff}$, but, of course, also by the exact Hamiltonian $\H_{\rot}$.  In fact, these Axioms are satisfied by an infinite number of Hamiltonians, so that more axioms are needed to define the effective Hamiltonian of the exact rotating wave approximation.  It has been surmised in Ref.~\cite{zeuch18} that a third axiom may be enough to distinguish the effective Hamiltonian from all other Hamiltonians that fulfill these two axioms.  Since the convergence of the series in Eq.~(\ref{HeffSeries}) is in general not guaranteed, we want the third axiom to be independent of this series definition.  

As motivated in the Introduction, the fact that the effective time evolution is in general significantly smoother than the exact time evolution implies that the effective trajectory traverses shorter paths for a given gauge parameter $\beta_0$.  Shorter paths translate to smaller (positive) eigenalues when averaged over the qubit trajectory and gauge parameter.  Hence our proposed third axiom can be formulated as follows,
\begin{axiom}
	There exists a functional of the form
	\begin{eqnarray}
		Q[\H(t; \beta_0)] &=& \int_{\beta_0 = 0}^{2\pi} \int_{\tau} f(\H(\tau; \beta_0)) \d \tau \d \beta_0,
		\label{axiom3}
	\end{eqnarray}
	where the integral over time $\tau$ extends over the duration of the pulse, and a certain integrand $f(\H(t; \beta_0))$, which is minimized by the effective Hamiltonian $\H_{\eff}$ given in Eq.~(\ref{HeffSeries}).
	\label{axiom3_axiom}
\end{axiom}
In the present study we scrutinize this third proposed axiom for two particular integrands introduced in the subsequent section.

\subsection{Proposed Integrands for the Third Axiom}
\label{integrands}

The main integrand $f$ considered by us is the positive eigenvalue of the Hamiltonian, $f_\I(\H(t; \beta_0)) = \eig_+(\H(t; \beta_0))$.  We also consider the positive eigenvalue of the derivative of the Hamiltonian, $f_\II(\H(t; \beta_0)) = \eig_+(\dot \H(t; \beta_0))$.  

In order to introduce integrands for the above functional (\ref{axiom3}), we begin by introducing a generic, traceless Hamiltonian $\H$ that is assumed to satisfy axioms 1 and 2 given above,
\begin{eqnarray}
	\H(t; \beta_0) = {\bf h}(t, \beta_0)\cdot \sigma = \lambda_+(t, \beta_0) \hat h(t, \beta_0) \cdot \sigma.
	\label{HGeneric}
\end{eqnarray}
Here we paramaterize the Hamiltonian using a three-dimensional vector ${\bf h} = \lambda_+ \hat h$ with positive eigenvalue $\lambda_+ = \lambda_+(t, \beta_0)$ and unit vector $\hat h = \hat h(t, \beta_0)$.  This Hamiltonian $\H$, given in Eq.~(\ref{HGeneric}), generates a stroboscopic time evolution as defined in Axiom \ref{axiom2}.  

Based on such a Hamiltonian, the first integrand proposed by us, denoted $f_{\I}$, is a measure for the size of the Hamiltonian given by its operator $2$-norm, or equivalently its positive eigenvalue,
\begin{eqnarray}
	f_{\I}(\H(t; \beta_0)) = \lVert \H(t; \beta_0) \rVert_2 = \eig_+(\H(t; \beta_0)) \equiv \lambda_+\H(t; \beta_0).
	\label{fIII_def}
\end{eqnarray}
The second integrand is a measure for the size of the time derivative of the Hamiltonian given by its operator $2$-norm, which similarly is equal to the positive eigenvalue of $\dot \H(t; \beta_0) = (\partial/\partial_t)\H(t; \beta_0)$,
\begin{eqnarray}
	f_{\II}(\H(t; \beta_0)) = \lVert \dot \H(t; \beta_0) \rVert_2 = \eig_+(\dot \H(t; \beta_0)).
	\label{fII_def}
\end{eqnarray}

The integrands $f_\I$ and $f_\II$ are, respectively, analytically analyzed in Secs.~\ref{minimizeEigenvalue} and \ref{directional}.


\subsection{Variational Minimization}
\label{minimization}

In this work we test the validity of Axiom \ref{axiom3_axiom} by numerically minimizing the functional (\ref{axiom3}).  To probe whether or not our effective Hamiltonian $\H_\eff$ constitutes a local minimum of this functional, we compute the integral for variational Hamiltonians $\H_\var$ in the vicinity of $\H_\eff$, i.e.
\begin{equation}
	\H_\var(t; \beta_0) = \H_\eff(t; \beta_0) + \delta \H(t; \beta_0)
	\label{vary_H}
\end{equation}
with $\lVert \delta \H\rVert \ll \lVert \H_\eff\rVert$ for some operator norm $\lVert \cdot \rVert$.  

When choosing such variational Hamiltonians $\H_\var$, we need to ensure that Axioms \ref{axiom1} and \ref{axiom2} are satisfied.  While analyticity (as required by Axiom \ref{axiom1}) is straightforward to built into this variational Hamiltonian, satisfying the requirement of stroboscopic time evolution stated in Axiom \ref{axiom2} is less trivial.  This is because the two effective time evolution operators $U_{\beta_0}(t, t_i) = \mathcal{T}e^{-i\int_{t_i}^{t} \d \tau \mathcal \H_\eff(\tau; \beta_0)}$ for some fixed initial time $t_i$ [cf.~Eq.~(\ref{Ueff})] and
\begin{equation}
	U_{\beta_0}^{\var}(t, t_i) = \mathcal{T}e^{-i\int_{t_i}^{t} \d \tau \mathcal H_\var(\tau; \beta_0)}
	\label{Uvar}
\end{equation}
with $\H_\var = \H_\eff + \delta \H$ as in Eq.~(\ref{vary_H}) for a time-dependent effective Hamiltonian cannot be related to one another straightforwardly.  This fact makes it difficult to find Hamiltonians $\delta H$ such that $\H_\var$ results in the same stroboscopic time evolution (\ref{goal0}) as $\H_\eff$.  

The most natural variation of the Hamiltonian as given in Eq.~(\ref{vary_H}) can be implemented by enforcing the condition of stroboscopic time evolution via Lagrange multipliers.  For this, the function (\ref{axiom3}) would be rewritten as
\begin{eqnarray}
	Q[\H(t; \beta_0)] &=& \int_{\beta_0 = 0}^{2\pi} \int_{\tau} f(\H(\tau; \beta_0)) \d \beta_0 \d \tau \nonumber \\ 
			&& \quad + \int_{\tau} \int_{\beta_0 = 0}^{2\pi} \mu(\H(\tau; \beta_0)) h(\tau, \beta_0) \lVert U_{\beta_0}(\tau, t_i) - U_{\exact}(\tau, t_i) \rVert \d \beta_0 \d \tau,
	\label{axiom3_lagrangified}
\end{eqnarray}
where again the $\tau$ integral extends over the entire duration of the pulse.  In the second integral, we have introduced a Lagrange multiplier $\mu(\H(t; \beta_0))$, and we use a function $h(t, \beta_0)$ that limits the two-dimensional integral to the stroboscopic points (\ref{sets}), $\{t_0, t_0 \pm t_c, t_0 \pm 2 t_c \ldots \}$ with $t_0 = \beta_0/(2\omega)$ and $t_c = \pi/\omega$.  For this let $h(t, \beta_0) = \sum_{n = -\infty}^\infty \delta(t - \beta_0/(2\omega) + n t_c)$; this ensures that for all $t$ and $\beta_0$ at which the effective and exact trajectories do \emph{not} need to coincide, the integrand of the second integral in Eq.~(\ref{axiom3_lagrangified}) is automatically zero.  The exact time evolution operator $U_{\exact}$ is given by Eq.~(\ref{Uexact}), so the factor $h(\tau, \beta_0)\lVert U_{\eff}(\tau, t_i) - U_{\exact}(\tau, t_i) \rVert$ constitutes the constraint of equal stroboscopic time evolution as stated in Axiom \ref{axiom2}.  

Below, however, we build Axiom \ref{axiom2} directly into the time evolution.  The basic idea is to vary the time evolution operator rather than the Hamiltonian.  We begin by writing the time evolution operator as a parameterization of a rotation around the Bloch sphere [referring to the last equality in Eq.~(\ref{Ugeneric})],
\begin{eqnarray}
	U_{\beta_0}(t, t_i) = e^{-i \overline{\H} t} = e^{-i {\bf n}(t, \beta_0) \cdot \sigma} = e^{-i \alpha(t, \beta_0) \hat n(t, \beta_0) \cdot \sigma}.
	\label{n_def}
\end{eqnarray}
That is, we parameterize the operator of the Magnus expansion as $\overline{\mathcal H}\, t = {\bf n} = \alpha \hat n$, where generally both $\alpha = \alpha(t, \beta_0)$ and $\hat n = \hat n(t, \beta_0)$ depend on both time $t$ and the gauge parameter $\beta_0$.  We note that the initial time $t_i$ is considered fixed, because of which we suppress the dependence on this initial time when writing the vector $\bf n(t, \beta_0)$.  

The next step is to introduce the variation
\begin{eqnarray}
	{\bf n}_\var(t, \beta_0) = {\bf n}(t, \beta_0) + \delta {\bf n}(t, \beta_0),
	\label{vary_n}
\end{eqnarray}
which replaces the direct variation of the Hamiltonian as given in Eq.~(\ref{vary_H}).  To ensure that the resulting variational time evolution operator (\ref{Uvar}), which at this point reads
\begin{equation}
	U_{\beta_0}^{\var}(t, t_i) = e^{-i {\bf n}_\var \cdot \sigma},
	\label{Uvar1}
\end{equation}
satisfies Axiom \ref{axiom2}, we choose functions $\delta {\bf n}(t, \beta_0)$ which are zero at the times $\{t_0, t_0 \pm t_c, t_0 \pm 2t_c, \ldots\}$ [cf.~Eq.~(\ref{sets})], i.e.,
\begin{equation}
	\delta {\bf n}(t = t_0 + n t_c; \beta_0) \equiv 0  \qquad \qquad \forall n \in \mathbb Z.
\end{equation}
Our explicit set of functions $\delta {\bf n}$ is given below in Sec.~\ref{trialFxns}.  

The integral of interest is then that given in Eq.~(\ref{axiom3}), in which we parametrize the Hamiltonian using the vector ${\bf n}$, $\H(t; \beta_0) = \H({\bf n}(t, \beta_0))$.  Our goal is thus to minimize the functional
\begin{eqnarray}
	Q[{\bf n}(t, \beta_0)] &=& \int_{\beta_0 = 0}^{2\pi} \int_{\tau} f( \H({\bf n}(\tau, \beta_0)) ) \d \tau \d \beta_0.  
	\label{axiom3n}
\end{eqnarray}

In Sec.~\ref{GaussianEnvelope} we present the envelope function used for the minimization.  In Sec.~\ref{symmetry} we then examine the symmetries of the various Hamiltonians; this symmetry consideration then allows us to consider a restricted set of trial functions, which reduces the complexity of the numerical minimization.



\subsubsection{Gaussian Envelope}
\label{GaussianEnvelope}

\begin{figure}
	\includegraphics[width = \columnwidth]{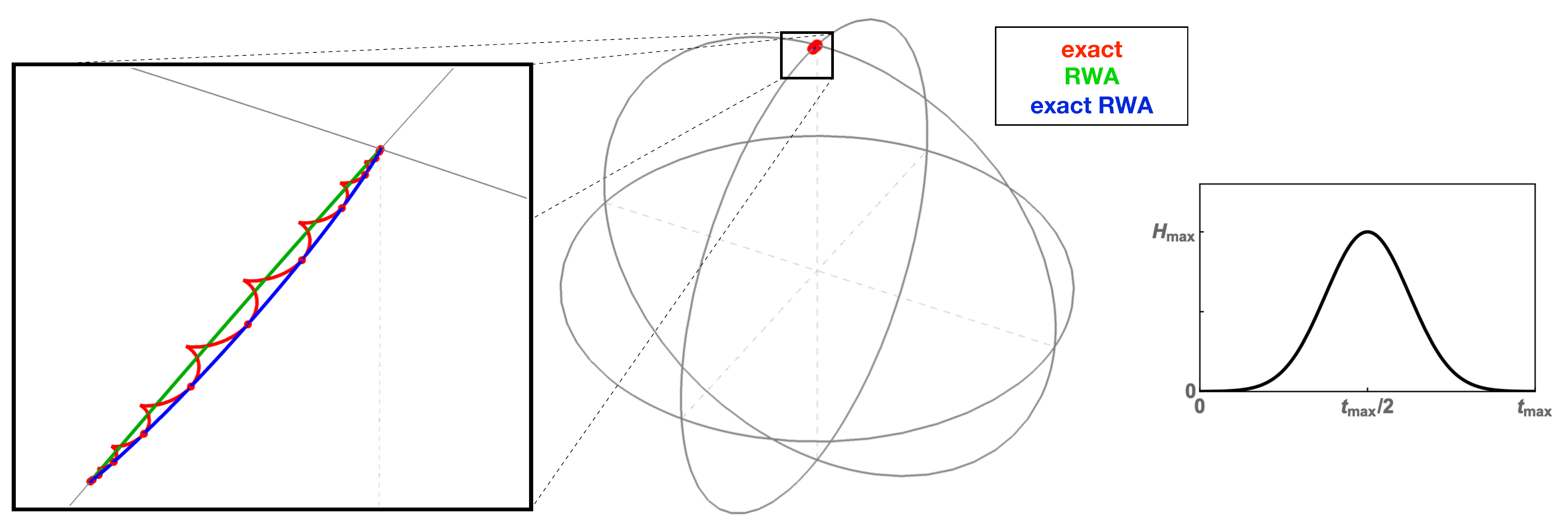}
	\caption{Actual rotation, similar to the rotations shown in Fig.~\ref{trajectories}, but that used in this study for the envelope (\ref{H1Gaussian}) with $H_{\max} = A = 0.002$  and $\sigma = 2t_c$.  The total angle traversed by the qubit vector is very small compared to the $\pi$ pulses of Fig.~\ref{trajectories}.}
	\label{actual_rotation}
\end{figure}

Our minimization method is based on a Gaussian envelope function of the form
\begin{eqnarray}
	H_1(t) &=& A e^{-\tfrac{t^2}{2\sigma^2}}, \qquad \qquad t\in [0, t_\gate].
	\label{H1Gaussian}
\end{eqnarray}
It is paramterized by an amplitude $A$ and a width $\sigma$.  

A convenient parameter range for $A$ and $\sigma$ is determined by the effective Hamiltonian series (\ref{HeffSeries}) to converge quickly.  This allows us to compute the effective time evolution in the exact rotating wave approximation to high precision while computing the effective Hamiltonian only up to moderate order in $1/\omega$.  To see which parameters lead to fast convergence, we take from the effective Hamiltonians given in Appendix \ref{effH_appendix} that we require $A/\omega$ to be small, which corresponds to weak driving.  This ensures that terms proportional to $H_1^{n+1}/\omega^{n}$ become negligible quickly.  Furthermore, we want terms propertional to the $n$th derivative $\dersub{H_1}{n}$ to fall off quickly with increasing $n$, which means that $\sigma \omega \sim \sigma/t_c$ (recall that $t_c = \pi/\omega$) should be large.  Besides, however, we want the fraction $\sigma/t_c$ to not be too large, so that the Gaussian function falls off sufficiently quickly.  This leads to a relatively short pulse, in the sense that it covers only a small number of Magnus intervals (\ref{MagnusInterval}) of duration $t_c$.  

This envelope function can be plotted for various parameters $A$ and $\sigma$ in the Mathematica notebook \texttt{exactRWA/programs/numerics/mathematica/Gaussian\_envelope.nb} available.  Based on this, we have made the choice for ou parameter values of $A = 0.002$, $\sigma = 2t_c = 4\pi$ with $\omega = 1/2$ (and thus $t_c = \pi/\omega = 2\pi$).  An exemplary time evolution due to this envelope function is shown in Fig.~\ref{actual_rotation}.  Based on these parameters, our effective Hamiltonian has been truncated only to order $1/\omega^5$, but given that $A/\omega$ is rather small the only terms that are appreciable in our calculation are those proportional to $A^3$.  This is an important feature for approximating  the effective time evolution for the driven qubit via the Magnus expansion, which is discussed below.



\subsubsection{Analytic approximation for effective time evolution}
\label{Ueff_analytic}

We approximate the effective time evolution for the Gaussian pulse using the Magnus expansion.  For this we consider the first three terms given explicitly in Appendix \ref{magnus_appendix}.  Furthermore, as noted above in the previous section, we only consider terms up to order $A^3$, which we are allowed to do since the ratio $A/\omega$ is very small.  

Here we approximate the time evolution analytically for the RWA Hamiltonian (\ref{HRWA}), which constitutes the lowest-order term in the effective Hamiltonian series (\ref{HeffSeries}), 
\begin{eqnarray}
	\H_{\RWA} = \frac{H_1(t)}4 \sigma_x \stackrel{(\ref{H1Gaussian})}{=}= \frac{A e^{-\tfrac{t^2}{2\sigma^2}}}4 \sigma_x.
\end{eqnarray}
To lowest order in the Magnus expansion of the time evolution operator , which is equivalent to ignoring time ordering completely of the time evolution operator, we then have
\begin{eqnarray}
	\overline \H^{(0)} &=& \frac1{t} \int_{t = \beta_0/(2\omega)}^{\beta_0/(2\omega) + t} \H_{\RWA}(t; \beta_0) \d t + \mathcal O(1/\omega) \nonumber \\
		&=& \frac{1}{4} \sqrt{\frac{\pi }{2}} \sigma \left(\text{erf}\left(\frac{t+\frac{\beta _0}{2\omega}}{\sqrt{2} \sigma }\right) - \text{erf}\left(\frac{\beta _0}{2 \sqrt{2} \sigma \omega}\right)\right) \sigma_x + \mathcal O(1/\omega).
	\label{ME_zero_RWA}
\end{eqnarray}
The full formulae for the analytic Magnus expansion used in our numerical minimization up to order $A^3$ can be found in the Mathematica notebook \texttt{MagnusExpansion\_results.nb}.\footnote{ \texttt{https://github.com/zeuch/exactRWA/programs/numerics/}}

\subsubsection{Symmetry}
\label{symmetry}

We use the symmetry of the envelope function to streamline the numerical minimization.  The envelope function of our choice, i.e., the Gaussian envelope discussed in the previous Sec.~\ref{GaussianEnvelope}, is an even function with respect to time reversal, $t \rightarrow -t$.  To see the implication on the symmetry of the Hamiltonian, first consider the \emph{generic} rotating-frame Hamiltonian (\ref{Hrot0}) for a symmetric envelope function $H_1(t)$ such as the Gaussian function described in the previous section, which fulfills $H_1(-t) = H_1(t)$.  Given this even symmetry, it is easy to see that the components of both $\sigma_x$ and $\sigma_z$ of this Hamiltonian are symmetric with respect to time reversal, while the $\sigma_y$-component is anti-symmetric.  

When comparing this symmetry property of the exact Hamiltonian to that of the effective Hamiltonian $\H_{\eff}(t; \beta_0)$, we need to take into account that the latter depends not only on time $t$ but also on the gauge parameter $\beta_0 = 2\omega t_0$ [cf.~Eq.~(\ref{beta0})].  Here $t_0$ defines the stroboscopic set of times (\ref{sets}), given by $\{t_0, t_0 \pm t_c, t_0 \pm 2t_c, \ldots\}$, at which the exact and effective time evolutions agree.  Note that time reversal maps the set (\ref{sets}) to $\{-t_0, -t_0 \pm t_c, -t_0 \pm 2t_c, \ldots\}$.  Given the proportionality relation between $t_0$ and $\beta_0$, the full symmetry operation for time reversal in case of the effective Hamiltonian $\H_\eff(t; \beta_0)$ is thus
\begin{eqnarray}
	S: \qquad (t, \beta_0) \ \rightarrow \ (-t, - \beta_0).
	\label{symmetryOp}
\end{eqnarray}
We find that, as above for the exact Hamiltonian, for this symmetry operation $S$ the $\sigma_x$- and $\sigma_z$-components of $\H_\eff$ are symmetric and the $\sigma_y$-component is anti-symmetric.  The exact and effective Hamiltonians thus have the same symmetry with respect to the operation $S$.

The Hamiltonian that minimizes the integral in Eq.~(\ref{axiom3}) may not need to have the same symmetry properties as the exact and effective Hamiltonians.  However, we restrict our search to Hamiltonians of that very symmetry described in the previous paragraph, since this allows us to reduce the size of the search space of variational Hamiltonians.  

Let us now determine the $S$-symmetry properties of the variational parameters $\delta {\bf n}(t; \beta_0)$, introduced above in Eq.~(\ref{vary_n}), for the assumption that the variational Hamiltonian fulfills the same symmetry as the exact and effective Hamiltonians.  To do this, consider a generic time evolution operator of the form (\ref{Ugeneric}) for $t_i = \beta_0$ and $t_f = \beta_0 + t$, or $U(t_f, t_i) = \mathcal{T}\exp(-i\int_{\beta_0}^{\beta_0 + t} \d \tau \mathcal H(\tau)) = e^{-i \overline{\mathcal H} t}$.  For simplicity, we focus on the lowest-order term of the Magnus expansion (\ref{MagnusSeries}) given in Appendix \ref{magnus_appendix}, $\overline{\mathcal H} \cong \overline{\mathcal H}^{(0)}$ [see also Eq.~(\ref{ME_zero_RWA})].  

We first note the following identity for the expression $\int_{\beta_0}^{\beta_0 + t} a(\tau, \beta_0) \d\tau$ with an arbitrary function $a(t, \beta_0)$ for the above symmetry operator $S$,
\begin{eqnarray}
	\int_{\beta_0}^{\beta_0 + t} a(\tau, \beta_0) \d\tau &\stackrel{S}{\rightarrow}& \int_{-\beta_0}^{-\beta_0 - t} a(\tau, -\beta_0) \d\tau \nonumber \\
	&\stackrel{\tau\rightarrow -\tau}=& \int_{\beta_0}^{\beta_0 + t} a(-\tau, -\beta_0) \d(-\tau) \nonumber \\
	&=& - \int_{\beta_0}^{\beta_0 + t} a(-\tau, -\beta_0) \d\tau.
	\label{symmetryIntegral0}
\end{eqnarray}
We now use this identity to find the $S$-symmetry for the $\bf n$ vector defined in Eq.~(\ref{n_def}) for the lowest-order Magnus expansion, ${\bf n}(t, \beta_0)\cdot \sigma = \overline {\H} \cong \overline {\H}^{(0)}$,
\begin{eqnarray}
	{\bf n}(t, \beta_0)\cdot \sigma = \int_{\beta_0}^{\beta_0 + t} \H(\tau) \d \tau &\stackrel{S,\  (\ref{symmetryIntegral0})}{\longrightarrow}& - \int_{\beta_0}^{\beta_0 + t} \H(-\tau) \d \tau.
	\label{symmetryIntegral}
\end{eqnarray}
We therefore consider the symmetry of the exact Hamiltonian (\ref{Hrot}) under time reversal,
\begin{eqnarray}
	\mathcal H_{\rot}(-t) &=& \frac{H_1(-t)}4 ( \sigma_x + \cos(-2\omega t) \sigma_x - ( \sin(-2\omega t)\sigma_y) ) \nonumber \\
		&=& \frac{H_1(t)}4 ( \sigma_x + \cos(2\omega t) \sigma_x - \sin(2\omega t)(-\sigma_y)),
	\label{symmetryH}
\end{eqnarray}
where we have used the fact that $H_1(t)$ is assumed to be symmetric.  Combining this result with Eq.~(\ref{symmetryIntegral}) we find that $n_x$ is odd and $n_y$ is even under the symmetry operation $S$ defined in Eq.~(\ref{symmetryOp}).  Using the more complete exact Hamiltonian (\ref{Hrot0}), one can similarly find that the $n_x$ is odd under this symmetry.  Recall that the effective and exact Hamiltonians have the same $S$-symmetry properties, because of which it suffices to consider one Hamiltonian.

\subsubsection{Trial Functions for Time Evolution Operator}
\label{trialFxns}

In our numerical variation we follow the convention
\begin{eqnarray}
	\omega = 1/2,
\end{eqnarray}
for which we have $t_c = \pi/\omega = 2\pi$.  Hence the Magnus interval (\ref{MagnusInterval}) are duration of $2\pi$,
\begin{eqnarray}
	[t_0 + n t_c, t_0 + (n+1) t_c) = [t_0 + 2\pi n, t_0 + 2\pi (n+1)).  
	\label{MagnusInterval_convention}
\end{eqnarray}

The central condition to be satisfied by the variational functions ${\bf n_{\var}} = {\bf n} + \delta {\bf n}$ [cf.~Eq.~(\ref{vary_n})] is that of stroboscopic as defined in Axiom \ref{axiom2}.  When varying this vector, this simply means that for the times (\ref{sets}) the variational vector ${\bf n_{\var}}(t; \beta_0)$ must coincide with the effective vector ${\bf n}(t; \beta_0)$, or 
\begin{eqnarray}
	\delta {\bf n}(t = t_0 + n t_c; \beta_0) = 0 \qquad\qquad \forall n \in \mathbb{Z}. 
	\label{zero}
\end{eqnarray}
We write the variational functions as a product
\begin{equation}
	\delta {\bf n}(t; \beta_0) = [\sin[(t-\beta_0)/2+\eta\phi]^2 - \sin(\eta\phi)^2] e^{- (1+c) t^2/(2 \sigma^2)} {\bf g}(t; \beta_0) = f_{\text{outer}}(t; \beta_0) {\bf g}(t; \beta_0)
	\label{trial_fxns}
\end{equation}
of a vector ${\bf g}(t; \beta_0)$ specified below, and an ``outer factor'' $f_{\text{outer}} = \sin[( t-\beta_0)/2 + \eta\phi]^2 - \sin(\eta\phi)^2$, which ensures that the condition (\ref{zero}) is fulfilled for arbitrary analytic ${\bf g}(t; \beta_0)$.  Furthermore, for a fixed value of $\beta_0$ this outer factor is nonzero on every entire Magnus interval (\ref{MagnusInterval_convention}), and the symmetry of this factor close to the boundary of the Magnus intervals can be varied by the variational parameter $\phi\in[0, 2\pi]$.  We note that we have also introduced two variational parameter $c\in\mathbb{R}$.  The factor $\eta \sim 10^6$, fixed for the a numerical minmization, attempts to reconcile the small incremental changes of the variational parameters compared to $\phi \in [0, 2\pi]$.


In principle, the ``starting vector'' $\n_0(t; \beta_0)$ [remove subscript:  replace with $\n(t; \beta_0)$] is either that of the effective or the exact Hamiltonian,
\begin{eqnarray}
	\n_0(t; \beta_0) =
	\begin{cases}
		\n_{\eff}(t; \beta_0)  \qquad \qquad \text{or}\\
		\n_{\exact}(t_0).
	\end{cases}
\end{eqnarray}
In our experiments we, of course, choose the effective ${\bf n}$ because that's the minimum.  

The ansatz for the function ${\bf g}(t; \beta_0) = (g_x, g_y, g_z)$ can suitably be written as a Fourier series.  First, we use the periodicity in $\beta_0$ to write
\begin{eqnarray}
		{\bf g}(t; \beta_0) = \sum_{m=0}^M {\bf A}_m(t) \cos(m \beta_0) + {\bf B}_m(t) \sin(m \beta_0),
	\label{ansatz1}
\end{eqnarray}
where ${\bf A}_m = (A_{x,m}, A_{y,m}, A_{z,m})$ and ${\bf B}_m = (B_{x,m}, B_{y,m}, B_{z,m})$.  Of course, Eq.~(\ref{ansatz1}) is periodic in $\beta_0$ with periodicity $2\pi$.  

For the temporal dependence we need to make sure that there is \emph{no} explicit periodicity of duration $2\pi$.  Generically for the component determined by $i=\{x, y, z\}$, we thus write
\begin{eqnarray}
	A_{i, m}(t) &=& \sum_{n=0}^{N} a_{i, m, n} \cos(nt/L) + a'_{i, m, n} \sin((n+1)t/L),
	\label{FourierTA} \\
	B_{i, m}(t) &=& \sum_{n=0}^{N} b_{i, m, n} \cos(nt/L) + b'_{i, m, n} \sin((n+1)t/L).
	\label{FourierTB}
\end{eqnarray}
These functions $A_{i, m}$ and $B_{i, m}$ have a period of $T_m = 2\pi/(m/L) = 2\pi L/m$, which should be different from the duration of a Magnus interval for completenes (otherwise we don't allow different Magnus intervals to be treated differently).  The value for $L$ that we found useful is $L = 5*\sigma$ (where $\sigma$ is the width of the Gaussian envelope).

The total number of parameters $\{a_{i, m,}, a'_{i, m,}, b_{i, m,}, b'_{i, m,}\}$ is then equal to $4\times3\times M \times N$ (3 for $i=x, y, z$).

\subsection{Kick Operators}
\label{kicks}

The kick operator formalism in relation to the exact rotating wave approximation was introduced in Ref.~\cite{zeuch18} in Sec.~3.  These operators are needed for drive envelope function $H_1(t)$ that are not entirely smooth, or in particular for for envelopes for which some derivative behaves like a $\delta$-function.  In this work we ignore the corrections due to kick operators, since they are so small that they do not affect our numerical results.

\subsection{Coding}
\label{coding}

The github page \texttt{https://github.com/zeuch/exactRWA.git} mentioned in the Introduction contains various Mathematica notebook and Python scripts that have been used to do the numerical minimization.  For example, as noted in Sec.~\ref{Ueff_analytic}, the explicit results of the Magnus expansion up to fifth order in $1/\omega$ for the case of a Gaussian envelope function can be found in \texttt{exactRWA/programs/numerics/MagnusExpansion\_results.nb}.  

The Python scripts used for our minimization can be found in \texttt{exactRWA/programs/numerics/python}.  


\section{Positive Eigenvalue of the Hamiltonian}
\label{minimizeEigenvalue}

Here we analyse the integrand $f_\I(\H) = \eig_+(\H)$, or the Hamiltonian's positive eigenvalue.  As noted in the Introduction, the Hamiltonian's eigenvalue is related to the length of the traversed trajectory on the Bloch sphere.  When comparing the lengths of the exact and effective trajectories (cf.~Fig.~\ref{trajectories}), the latter are significantly shorter---suggesting that the integral over the total pulse duration of the Hamiltonian's positive eigenvalue may be minimized by the effective Hamiltonian.  

As described in Sec.~\ref{minimization}, we introduce the vector ${\bf n} = \alpha \hat n$ for the parameterization of the time evolution operator, $U = e^{-i {\bf n} \cdot \sigma}$ [as given in Eq.~(\ref{n_def})].  We use the Magnus expansion to compute this vector ${\bf n}(t, \beta_0) = \alpha(t, \beta_0) \hat n(t, \beta_0)$, where $\alpha(t, \beta_0)$ is a scalar and $\hat n(t, \beta_0)$ is a three-dimensional unit vector.  In order to carry out the variational minimization of the functional $Q[\bf {n}]$ as given in Eq.~(\ref{axiom3n}) for the integrand $f_\I$, we need to relate this integrand to the vector $\bf {n}$.  To do this, we first write down the Schroedinger equation for the time evolution operator,
\begin{eqnarray}
	\frac{\partial U(t, t_i)}{\partial t} = -i \H(t; \beta_0) U(t, t_i),
	\label{schroedingerU}
\end{eqnarray}
which allows us to rewrite the positive eigenvalue of the Hamiltonian as follows,
\begin{eqnarray}
	f_\I &=& \lambda_+(t, \beta_0) = \eig _+\left(\H\right)
			= \eig _+\left[\left(i \frac{\partial U}{\partial t}\right) U^{\dagger}\right].
	\label{f1_of_U}
\end{eqnarray}
In principle, one could thus use this equation to compute the integrand $f_\I$ for a given $\bf {n}$ by first determining the time evolution operator.  However, since we directly vary this $\bf {n}$ vector in our numerical minimization it is advantageous to have a direct algebraic relation between $f_\I$ and the $\bf {n}$ vector, which is computed below.

We note that one needs to take some care when computing the temporal derivative $\frac{\partial U}{\partial t}$, which is required to evaluate Eq.~(\ref{f1_of_U}), if the time evolution operator is parameterized as $U = \exp(f(t))$ with an operator $f(t)=-i{\bf n}\cdot \sigma$ [as opposed to the usual representation using the time ordering parameter].  This calculation is nontrivial if the operator $f(t)$ does not commute with its derivative, or $[f(t), \dot f(t)] \neq 0$; one way to see this is by noting that
\begin{eqnarray}
	\partial_t \exp(f) &=& \partial_t  (1 + f + \frac{f^2}{2!} + \frac{f^3}{3!} + \ldots) = \dot f + \frac{\dot f f + f \dot f}{2!} + \frac{\dot f f^2 + f \dot f f + f^2 \dot f}{3!} + \ldots
	\label{nontrivial}
\end{eqnarray}
cannot be regrouped straightforwardly as a power series.  The calculation of this derivative is carried out starting with a relatively generic function $f(t)$ in Appendix \ref{Udot}.  As discussed below in Sec.~\ref{t-dependent}, in the case of an SU(2) operator $f(t)$, this derivative can be obtained rather straightforwardly.  

In the remainder of this section we determine the integrand $f_I$ as a function of the ${\bf n}$ vector by evaluating Eq.~(\ref{f1_of_U}).  This is done for the two different cases of a constant drive amplitude $H_1(t) = H_1$ [Sec.~\ref{t-independent}] and a time-dependent $H_1(t)$ [Sec.~\ref{t-dependent}].  In the former case, Eq.~(\ref{f1_of_U}) can be simplified straightforwardly, and even the functional (\ref{axiom3n}) can be evaluated completely analytically.  In the latter case, the result for $f_\I$ is significantly less trivial, because of which we minimize Eq.~(\ref{axiom3n}) numerically.  


\subsection{Constant Drive}
\label{t-independent}

Consider the case of a constant envelope function.  As explained in the Introduction [cf.~the discussion leading to Eq.~(\ref{Ueff_const})], the effective Hamiltonian for this case is itself a constant, or
\begin{equation}
	\qquad \qquad \qquad \qquad \H_\eff(t; \beta_0) = \H_\eff(\beta_0),	\qquad \qquad (H_1(t) \equiv H_1).
\end{equation}
The simplified time evolution operator (\ref{Ueff_const}) is given by
\begin{eqnarray}
	U_{\beta_0}(t, t_i) = e^{-i \H_{\eff}(\beta_0) (t-t_i)},
	\label{Ueff_constant}
\end{eqnarray}
so that for the parametrization (\ref{n_def}), $U = e^{-i {\bf n} \cdot \sigma}$, we find that the vector ${\bf n} = \alpha {\hat n}$ factors into a time-independent unit vector ${\hat n}(t, \beta_0) = {\hat n}(\beta_0)$ and a factor 
\begin{equation}
	\alpha(t, \beta_0) = c(\beta_0)  (t-t_i),
	\label{alpha_constant}
\end{equation}
which depends linearly on time.  

Let us now analyse the $\beta_0$-dependence of $\alpha(t, \beta_0)$, which is contained in the factor $c(\beta_0)$.  Given that the Hamiltonian is a sum of Pauli matrices [recall that, as noted in the Introduction (cf.~Footnote \ref{foot:traceless}) we consider traceless Hamiltonians], we note for a unitary matrix $\H_{\eff} = {\bf n}\cdot \sigma = n_1\sigma_x + n_2\sigma_y + n_3\sigma_z$ with real parameters $n_1$, $n_2$ and $n_3$,
\begin{eqnarray}
	\eig_+(\H_{\eff}) = \eig_+(n_1\sigma_x + n_2\sigma_y + n_1\sigma_z) = \sqrt{n_1^2+n_2^2+n_3^2} = \lVert \H_{\eff} \rVert_2,
	\label{eigenvalues}
\end{eqnarray}
with the 2-norm $\lVert \cdot \rVert_2$.  Equation (\ref{alpha_constant}) thus becomes
\begin{eqnarray}
	\alpha(t, \beta_0) = \lVert \H_{\eff}(\beta_0) \rVert_2 (t-t_i).
	\label{alpha_constant_prime}
\end{eqnarray}
It follows from the periodicity of the exact Hamiltonian for a constant envelope that all different effective $\beta_0$-trajectories are traversed with one and the same angular rotation velocity.  Because of this, the norm of the Hamiltonian cannot depend on $\beta_0$, and we have
\begin{eqnarray}
	c(\beta_0) = \lVert \H_{\eff}(\beta_0) \rVert_2 \equiv \tilde c = \text{constant} > 0.
	\label{c}
\end{eqnarray}

Now consider the functional of interest (\ref{axiom3n}), $Q = \int_{\beta_0} \int_{\tau} f_\I \d \tau \d \beta_0$.  For this we need to compute the integrand (\ref{f1_of_U}).  To do this, we note that the derivative of the time evolution operator $U$ as parameterized in Eq.~(\ref{Ueff_constant}) is simple to compute.  This is because the difficulty described above in Eq.~(\ref{nontrivial}) does not occur, since the Hamiltonian is time-independent.  We have\footnote{Since the time evolution operator is, in this case, defined without the time ordering operator, this result also follows directly from the Schroedinger equation (\ref{schroedingerU}).}
\begin{eqnarray}
	\partial_t U_{\beta_0}(t, t_i) = -i\H_{\eff}(\beta_0)U_{\beta_0}(t, t_i).
	\label{U_dot}
\end{eqnarray}
This implies for the integrand $f_\I$ taken for the effective time evolution,
\begin{eqnarray}
	f_\I( \H_{\eff}({\bf n}(\tau, \beta_0)) ) &=& \eig_+\left[\left(i \frac{\partial U_{\beta_0}}{\partial t}\right) U^{\dagger}\right] \nonumber \\
			&\stackrel{(\ref{U_dot})}{=}& \eig_+\left[  \H_{\eff}U_{\beta_0} U_{\beta_0}^{\dagger}\right] \nonumber \\
			&\stackrel{(\ref{eigenvalues})}=& \lVert \H_{\eff} \rVert_2 \nonumber \\
			&\stackrel{(\ref{c})}=& |\tilde c| \label{f_I-useful_now} \\
			&\stackrel{(\ref{alpha_constant_prime})}=& |\dot \alpha(t, \beta_0)|. 
			\label{f_I-useful_later}
\end{eqnarray}
The functional $Q$ can thus be evaluated via
\begin{eqnarray}
	Q[{\bf n}(t, \beta_0)] &\stackrel{(\ref{f_I-useful_now})}=& \int_{\beta_0 = 0}^{2\pi} \int_{\tau=t_i}^{t_i+n t_c}  \tilde c \d \tau \d \beta_0 \nonumber \\
		&=& 2\pi n t_c \int_{\beta_0 = 0}^{2\pi} \tilde c.
	\label{axiom3constant}
\end{eqnarray}

This result means there is (probably) a degeneracy in the functional $Q$ for a constant drive envelope, which includes our effective Hamiltonian.  Computing the same integral $Q$ numerically for the exact time evolution results in a larger value, $Q_\exact > Q_\eff$, which shows that this degeneracy does not include the exact Hamiltonian.

\subsection{Generic (Analytic) Drive}
\label{t-dependent}

\subsubsection{Derivative of Time Evolution Operator}

As noted above, in Appendix \ref{Udot} we note a technique for computing the time derivate of a time evolution operator of the form $e^{-i f(t)}$, where $f(t)$ is an operator that does not commute with its derivative.  Thanks to a hint given to us by Alwin van Steensel, we are able to use an easier way of computing the derivative of the time evolution operator (\ref{n_def}), $U_{\beta_0}(t, t_i) = e^{-i {\bf n}(t, \beta_0) \cdot \sigma}$.  To do this, we again first write ${\bf n} = \alpha \hat n$, so that
\begin{eqnarray}
	\partial_t U(t)
		&=& \partial_t [\cos(\alpha) - i \hat n \cdot \sigma \sin\alpha] \nonumber\\
		&=& \dot \alpha \sin\alpha - i \dot {\hat n} \cdot \sigma \sin\alpha - i \dot \alpha \hat n \cdot \sigma \cos\alpha \nonumber\\
		&=& - i \dot {\hat n} \cdot \sigma \sin\alpha - i \dot \alpha \hat n \cdot \sigma [ \cos\alpha - i \hat n \cdot \sigma \sin\alpha ] \nonumber\\
		&=& - i( \dot {\hat n} \cdot \sigma \sin\alpha +  \dot \alpha \ \hat n \cdot \sigma \underbrace{e^{-i\alpha \hat n \cdot \sigma }}_{=U(t)}).
\end{eqnarray}
Using this, our term in the integrand (\ref{f1_of_U}) can be calculated as follows,
\begin{eqnarray}
	i(\partial_t U) U^{\dagger}
		&=& \dot {\hat n} \cdot \sigma \sin\alpha \underbrace{[\cos\alpha + i \hat n \cdot \sigma \sin\alpha]}_{ = e^{i\alpha \hat n\cdot \sigma} = U^\dagger} +  \dot \alpha \ \hat n \cdot \sigma UU^{\dagger} \nonumber\\
		&\stackrel{(\ref{sigmaCross})}{=}&  \dot {\hat n} \cdot \sigma \sin\alpha \cos\alpha - |\dot {\hat n}|(\hat n \times \frac{\dot {\hat n}}{|\dot {\hat n}|}) \cdot \sigma \sin\alpha^2 + \dot \alpha \hat n \cdot \sigma  \nonumber\\
		&=& (1/2) \sin (2\alpha) \dot{\hat n}\cdot \sigma  - (1/2) [1-\cos (2\alpha)] (\hat n \times \dot {\hat n}) \cdot \sigma +  \dot \alpha \hat n \cdot \sigma 
		\label{AlwinUseful}\\
		&=& (1/2)|\dot {\hat n}| \sin (2\alpha) \frac{\dot {\hat n}}{|\dot {\hat n}|}\cdot \sigma  - (1/2) |\dot {\hat n}| [1-\cos (2\alpha)]  (\hat n \times \frac{\dot {\hat n}}{|\dot {\hat n}|}) \cdot \sigma +  \dot \alpha \hat n \cdot \sigma ,
		\label{revisited}
\end{eqnarray}
where we used
\begin{eqnarray}
	(a\cdot \sigma)(b\cdot \sigma) &=& a\cdot b + i (a\times b) \cdot \sigma, \label{sigmaCross} \\
	\sin(x)\cos(x)  &=& \frac12\sin(2x), \\
	\sin(x)^2 		&=& \frac12 (1-\cos(2x)).
\end{eqnarray}
We have also used the following simplifying step in computing the eigenvalue,
\begin{eqnarray}
	\sin(\alpha)^2 + (\cos(\alpha) - 1)^2 = \sin(\alpha)^2 + \cos(\alpha)^2 -2 \cos(\alpha) + 1 = 2(1-\cos\alpha).
\end{eqnarray}

The integrand $f_{\I}$ can be computed directly from the equation (\ref{revisited}).  Denoting $\hat d_1 = \hat n$, $\hat d_2 = \hat {\dot {\hat n}}$ and $\hat d_3 = \hat n \times \hat {\dot {\hat n}}$, we can write
\begin{eqnarray}
	\H &=& d_1 \hat d_1\cdot \sigma + d_2 \hat d_2\cdot \sigma + d_3 (\hat d_1 \times \hat d_2)\cdot\sigma \\
	\Rightarrow f_{\I} &=& \eig_+ (\H) = \sqrt{d_1^2 + d_2^2 + d_3^2}.
	\label{posEigV}
\end{eqnarray}
Using this, we find the result
\begin{eqnarray}
	f_\I( \H_{\eff}({\bf n}) ) &=& \eig_+\left[\left(i \frac{\partial U_{\beta_0}}{\partial t}\right) U_{\beta_0}^{\dagger}\right] 
			\stackrel{(\ref{revisited})}{=} \sqrt{\dot \alpha^2 + (1/2)|\dot {\hat n}|^2 [1 - \cos (2\alpha)]}.
	\label{Qf_I-Final}
\end{eqnarray}

Note that in the case of a constant drive envelope the vector $\hat n$ (the rotation axis of the effective evolution) is a constant.  Since this implies $\dot {\hat n} = 0$, this result (\ref{Qf_I-Final}) then reduces to the integrand for constant drive amplitudes computed above in Eq.~(\ref{f_I-useful_later}).




\section{Minimize the Variation of the Hamiltonian's Directional Vector}
\label{directional}

Here we follow the idea that the effective Hamiltonian has no fast-oscillating terms (no terms that oscillate with frequency $\omega$).  This is, of course, a striking difference between the exact Hamiltonian [see, e.g., Eq.~(\ref{Hrot})] and the effective Hamiltonian [see, e.g., Eq.~(\ref{Heff_main})].  

In symbols, we want to minimize the variations of the vector $\hat h$, which is a measure for the noncommutativity of the Hamiltonian (\ref{HGeneric}).  As usual, the minimization (\ref{axiom3n}) is over both time $t$ and gauge parameter $\beta_0$ on the domain of an entire pulse.  When writing the Hamiltonian as in Eq.~(\ref{HGeneric}),
\begin{eqnarray}
	\H(t; \beta_0) = {\bf h}(t,\beta_0) \cdot \sigma,  \qquad \text{ where }{\bf h} = (h_x, h_y, h_z)^T,
	\label{HVia_h}
\end{eqnarray}
we define the integrand $\lambda_\II$ as
\begin{eqnarray}
	f_{\II}(t, \beta_0) = |\partial_t h| = \sqrt{(\partial_t h_x)^2 + (\partial_t h_y)^2 + (\partial_t h_z)^2}.
	\label{II--fxnOfH}
\end{eqnarray}
This can also be viewed as the matrix norm (more precisely, the 2-norm) of the time derivative of the Hamiltonian,
\begin{eqnarray}
	\lambda_ \II = \lVert \dot {\bf h}(t,\beta_0) \cdot \sigma\rVert_2 = \lVert \dot \H\rVert_2 \equiv \eig_+(\partial_t \H),
\end{eqnarray}
which is also equal to the positive eigenvalue of the Hamiltonian.  

For our minimization, we want to compute the integrand $\lambda_\II$ as a function of the $\bf n$ vector.  To do so, we again use the Schroedinger equation for the time evolution operator given in Eq.~(\ref{schroedingerU}),
\begin{eqnarray}
	f_{\II} &=& \eig_+(\partial_t \H) \nonumber \\
		&\stackrel{(\ref{schroedingerU})}{=}& \eig_+\left\{\partial_t \left[\left(i \frac{\partial U}{\partial t}\right) U^{\dagger}\right] \right\}.
	\label{easyII}
\end{eqnarray}

\subsection{Constant Drive Envelope}

For constant drive envelopes, recall that we have computed above in Sec.~\ref{t-independent} the positive eigenvalue of the Hamiltonian as a function of the ${\bf n}$ vector.  In that calculation we noted that the time evolution operator for constant $H_1(t) \equiv H_1$ can be written as in Eq.~(\ref{Ueff_constant}), or $U_{\beta_0}(t, t_i) = e^{-i\H_{\eff}(\beta_0) (t-t_i)}$.  As noted above in Eq.~(\ref{U_dot}), since the operator inside the exponential is self-commutative at arbitrary different times, we can simply write
\begin{eqnarray}
	f_{\II} = \eig_+ \left\{\partial_t\left(i(\partial_t U_{\beta_0})U^\dagger_{\beta_0}\right) \right)\} = \eig_+ \{ \partial_t \H_{\eff} \}.
\end{eqnarray}
That is, for the effective Hamiltonian $\H_{\eff}$, which for constant drive envelopes is independent of time $t$ so that $\partial_t \H_{\eff} = 0$, the integrand is zero,
\begin{eqnarray}
	f_{\II} = 0 \qquad\qquad\qquad (\H_{\eff}, H_1(t) = H_1).
\end{eqnarray}
In this case the integral in Eq.~(\ref{axiom3n}) is also zero, which suggests that here the effective Hamiltonian does indeed satisfy Axiom \ref{axiom3_axiom}.

\subsection{Generic (Analytic) Drive}
\label{explicit_formulaII}

We now use the intermediate step (\ref{revisited}) in the computation of the integrand $\lambda_\I = \eig_+(\H)$ given above to simplify the Hamiltonian $\H = {\bf h}\cdot \sigma$ appearing in Eq.~(\ref{easyII}),
\begin{eqnarray}
	\H &=& i(\partial_t U) U^{\dagger} \nonumber \\
			&\stackrel{(\ref{revisited})}{=}&(1/2) \sin (2\alpha) \dot{\hat n}\cdot \sigma  - (1/2) [1-\cos (2\alpha)] (\hat n \times \dot {\hat n}) \cdot \sigma +  \dot \alpha \hat n \cdot \sigma
	\label{incomplete} \\
		&=& \left( (1/2) \sin (2\alpha) \dot{\hat n} - (1/2) [1-\cos (2\alpha)] (\hat n \times \dot {\hat n}) \sigma +  \dot \alpha \hat n \right) \cdot \sigma
		\label{H_of_n_ThanksAlwin}
\end{eqnarray}
Above in Sec.~\ref{minimizeEigenvalue}, we have used the fact that this Hamiltonian is written as a sum of three perpendicular vectors $\hat d_1 = \hat n$, $\hat d_2 = \hat {\dot {\hat n}}$ and $\hat d_3 = \hat n \times \hat {\dot {\hat n}}$, which has allowed us to use Eq.~(\ref{posEigV}) to find an algebraic expression for $f_{\I}$.  

In order to obtain a similar expression for $f_{\II}$, we now compute the temporal derivative of the Hamiltonian $\H$ as given in Eq.(\ref{incomplete}),
\begin{eqnarray}
	\partial_t  \H &=& \partial_t \left[i(\partial_t U) U^{\dagger} \right] \nonumber \\
			&=& (1/2) \sin (2\alpha) \ddot{\hat n}\cdot \sigma + \dot \alpha \cos (2\alpha) \dot{\hat n}\cdot \sigma  - (1/2) [1-\cos (2\alpha)] (\hat n \times \ddot {\hat n}) \cdot \sigma \nonumber \\
				&& \qquad \qquad \qquad \qquad - \dot \alpha \sin(2\alpha) (\hat n \times \dot {\hat n}) \cdot \sigma + \dot \alpha \dot {\hat n} \cdot \sigma + \ddot \alpha \hat n \cdot \sigma \nonumber \\
			&=& (1/2) \sin(2\alpha) \ddot{\hat n}\cdot \sigma - (1/2) [1-\cos (2\alpha)] (\hat n \times \ddot {\hat n}) \cdot \sigma + \dot \alpha [\cos (2\alpha)+1] \dot{\hat n}\cdot \sigma   \nonumber \\
				&& \qquad \qquad \qquad \qquad - \dot \alpha \sin(2\alpha) (\hat n \times \dot {\hat n}) \cdot \sigma + \ddot \alpha \hat n \cdot \sigma
	\label{revisited2}\\
			&=& (1/2) \sin(2\alpha) \ddot{\hat n}\cdot \sigma - (1/2) [1-\cos (2\alpha)] (\hat n \times \ddot {\hat n}) \cdot \sigma + \dot \alpha n_v [\cos (2\alpha)+1] \hat n_v \cdot \sigma   \nonumber \\
				&& \qquad \qquad \qquad \qquad - \dot \alpha n_v \sin(2\alpha) \hat n_\perp \cdot \sigma + \ddot \alpha \hat n \cdot \sigma.
	\label{revisitedFinal}
\end{eqnarray}
In the last line we used $\dot {\hat n} = n_v \hat n_v$, and $(\hat n \times \dot {\hat n}) = n_v (\hat n \times {\hat n}_v) = n_v \hat n_\perp$ [$\hat n_\perp = \hat n_v \times \hat n$, see also below in Eq.~(\ref{theVectors})].  Just to be clear, with Eq.~(\ref{easyII}) it is clear that the integrand $f_{\II}$ is related to Eq.~(\ref{revisited2}) in that it is the (positive) eigenvalue, that is, $f_{\II} = \eig_+\left\{\partial_t \left[i(\partial_t U) U^{\dagger} \right]\right\}$.  

Considering that a unit vector $\hat n$ is perpendicular to its first derivative,
\begin{eqnarray}
	\dot{\hat n} \perp \hat n,
\end{eqnarray}
we can ``of course'' write the derivative of $\dot {\hat n} \equiv \vec n_v$ (the velocity vector of $\hat n$) as a sum of terms that are parallel-to and orthogonal-to $\hat v$.  This is what we did in Appendix \ref{unit_vectors_elegant}.  We need only write this new velocity vector as $\vec n_v = |n_v| \hat n_v$.  The three orthogonal vectors are then
\begin{eqnarray}
	\hat {\tilde x} \equiv \hat n, \qquad \hat {\tilde y} \equiv \hat n_v \equiv \frac{1}{|\dot {\hat n}|}\dot {\hat n}, \qquad \hat {\tilde z} \equiv \hat n_\perp = \hat n_v\times \hat n.
	\label{theVectors}
\end{eqnarray}
Since these vectors $\{\hat {\tilde x}, \hat {\tilde y}, \hat {\tilde z}\}$ form a right-handed coordinate system it follows, for instance, that
\begin{eqnarray}
	\hat {\tilde x} \times \hat {\tilde z} = - \hat {\tilde y}.
\end{eqnarray}

We can probably compute an algebraic equation for this integrand using Eq.~(\ref{revisitedFinal}) and the results computed in Appendix \ref{unit_vectors_elegant}.  The result is that the second derivative of the unit vector $\hat n$ is that given in Eq.~(\ref{ddotnHatApp})
\begin{eqnarray}
	\ddot {\hat n} =\dot n_v \hat n_v + n_v \left[(\vec n_a \cdot \hat n) \ \hat n + (\vec n_a \cdot \hat n_\perp) \ \hat n_{\perp} \right],
	\label{ddotnHat}
\end{eqnarray}
where 
\begin{eqnarray}
	\hat n_{\perp} &=& \hat n \times \hat n_v, \\
	\vec n_a &=& \dot {\hat n}_v.
\end{eqnarray}

Note that since $\vec n_a$ is the derivative of a unit vector (${\hat n}_v$), its dimension is [1/time] even though it plays the role of an acceleration vector.

\subsubsection{Closed-form Expression for $f_{\II}$}

The expression for the exact result of $f_{\II}$ is computed in the Mathematica notebook \texttt{integrand2.nb}, which can be found in the github repository [cf.~Sec.~\ref{coding}] in the folder \texttt{/exactRWA/programs/variational\_minimization}.  Combining Eqs.~(\ref{revisitedFinal}) and (\ref{ddotnHat}), with $f_{\II} = \eig_+(\partial_t \H)$ we have
\begin{eqnarray}
	f_{\II}^2 = \sin(\alpha)^2 [n_v(\vec n_a\cdot \hat n_{\perp} - 2\dot \alpha) \cos\alpha - \dot n_v \sin \alpha]^2 + [\ddot \alpha + n_v \vec n_a \cdot \hat n \cos \alpha \sin\alpha ]^2  \nonumber \\
			+ \frac14 [n_v (\vec n_a \cdot \hat n_\perp + 2\dot \alpha) - n_v (\vec n_a\cdot \hat n_\perp - 2\dot \alpha)\cos(2\alpha) + \dot n_v \sin(2\alpha) ]^2.
\end{eqnarray}

For coding this integrand, it is better if we rewrite the derivative of the Hamiltonian (\ref{H_of_n_ThanksAlwin}), which is given in Eq.~(\ref{revisited2}).  We focus on its defining vector, $\dot \H = \dot {\bf h}\cdot \boldsymbol{\sigma}$, like this,
\begin{eqnarray}
	\dot {\bf h} = a \ddot{\hat n} - b (\hat n \times \ddot {\hat n}) + c \hat n_v - d \hat n_\perp  + \ddot \alpha \hat n,
	\label{hDot}
\end{eqnarray}
where
\begin{eqnarray}
    a &=& (1/2) \sin(2\alpha),  \qquad b = (1/2) [1-\cos (2\alpha)],  \qquad c = \dot \alpha n_v [\cos (2\alpha)+1], \nonumber \\
     \ \ \ \ d &=& \dot \alpha n_v \sin(2\alpha), \qquad \text{and } n_v \stackrel{(\ref{theVectors})}{=} |\dot {\hat n}|.
\end{eqnarray}
Given the following orthogonality relations: $\ddot {\hat n} \perp (\hat n \times \ddot {\hat n})$, $\hat n \perp (\hat n \times \ddot {\hat n})$, and $\{\hat n, \dot {\hat n}, \hat n_\perp\}$ are pairwise orthogonal, we find
\begin{eqnarray}
	\dot {\bf h} \cdot \dot {\bf h}&=& a^2 \ddot{\hat n}\cdot\ddot{\hat n} + b^2 (\hat n \times \ddot {\hat n})\cdot(\hat n \times \ddot {\hat n}) + c^2 + d^2 + \ddot \alpha^2 \nonumber \\
	    && + 2 (ac \ddot{\hat n}\cdot\hat n_v - ad \ddot{\hat n}\cdot \hat n_\perp + a\ddot \alpha \ddot{\hat n}\cdot\hat n - bc (\hat n \times \ddot {\hat n})\cdot\hat n_v + bd(\hat n \times \ddot {\hat n})\cdot \hat n_\perp) \nonumber \\
		&=& a^2 |\ddot{\hat n}|^2 + b^2 |(\hat n \times \ddot {\hat n})|^2 + c^2 + d^2 + \ddot \alpha^2 \nonumber \\
	    && + 2 (ac \ddot{\hat n}\cdot\hat n_v - ad \ddot{\hat n}\cdot \hat n_\perp + a\ddot \alpha \ddot{\hat n}\cdot\hat n - bc (\hat n \times \ddot {\hat n})\cdot\hat n_v + bd(\hat n \times \ddot {\hat n})\cdot \hat n_\perp).
	\label{hDotSquared}
\end{eqnarray}


\section{Results of Integrals}
\label{Results_of_Integrals}

In Secs.~\ref{integrand1} and \ref{integrand1_full} we give results for the integrals that were found by our minimization.  Sections \ref{errorSources} and \ref{errorAnalysis} present our error analysis, and, most importantly, Sec.~\ref{caseAgainstI} gives the main result of this study:  the comparison of the integral results for the effective and variational Hamiltonians in relation to the maximal error.  This comparison gives a strong argument that our proposed set of axioms is incorrect.  

We have run an extensive numerical minimzation of the functional (\ref{axiom3n}).  In this numerical part of our work, we implemented the gradient descent algorithm called the Broyden–Fletcher–Goldfarb–Shanno algorithm\footnote{\url{https://en.wikipedia.org/wiki/Broyden\%E2\%80\%93Fletcher\%E2\%80\%93Goldfarb\%E2\%80\%93Shanno\_algorithm}}, in which the gradient is approximated.  We have compared the performance with the Nelder-Mead algorithm\footnote{\url{https://en.wikipedia.org/wiki/Nelder\%E2\%80\%93Mead\_method}}, which gave similar results.  Some integral results for $Q$ and uncertainties for the quantities $\alpha$ and $\bf n$, which are obtained numerically, are documented in Table \ref{integral_results}.

\begin{table}[t]
\begin{tabular}{|c||c|}\hline
	integrand 			& ${\bf n}_{\eff}$ $\mathcal{O}\left(A^3\right)$ \\ \hline\hline
	$f_{\text{I}}$(*)	 & 0.099116584(3) \\ \hline
	$f_{\text{II}}$ 	 & 0.00633432(1) \\\hline
	$\delta \alpha$ & $1.84\times10^{-8}$ \\
	$\delta n$ 		& $(1.84\times10^{-8}, 1.74\times10^{-8}, 1.09\times10^{-8})$ \\
	$\delta \hat n$, $|\delta \hat n|$ & $(7.64\times10^{-8}, 1.49\times10^{-6}, 9.61\times10^{-7})$, $1.50\times10^{-6}$ \\ \hline
	$\delta Q$ 		& $3.4(1)\times 10^{-6}$ [very conservative (constant maximum errors)] \\
	$\delta Q$ 		& $2.90 \times 10^{-6}$ [quadrature (constant maximum errors)] \\\hline
\end{tabular}
\caption{Specific results of integrals $Q$ for amplitude $A = 0.002$.  The number in parentheses next to the integrand denotes the upper bound in the number of trial functions $M$ [we use $M = N$, compare Eqs.~(\ref{ansatz1}) and (\ref{FourierTA})].  The number in [square brackets] denotes the numerical run that yields the results.  The star (*) indicates that we have confirmed that $f_{\I}$ and $f_{\I}^{\text{simplified}}$ yield the same results to the given accuracy.  The quantities $\delta n$,  $\delta \alpha$, $\delta \hat n$ and $|\delta \hat n|$ are absolute (not relative) errors.  }
\label{integral_results}
\end{table}

\subsection{Integrand 1}
\label{integrand1}


We improved the ${\bf n}$ vector to higher order, namely it is now proportional to $A^2$---below we denote it $\hat n_{A^2}$, and the old vector is called ${\bf n}_A$.  Note that the new vector ${\bf n}_{A^2}$ also goes to higher order in $1/\omega$.  

We wanted to find out if this improved ${\bf n}$ vector yields the same lower variational minimum for $f_{\I}$.  To do this, we first tried to find out if we could even repeat the old calculation with ${\bf n}_A$.  The results for this $f_{\I}$ integral is recorded in ``workMac440.txt'' (the original calculation was done in ``workMac44.txt").  That is, in ``440'' we redid the calculation with ${\bf n}_A$ just to be sure we can still do it with my current code.  Note that (it looks like) in recalculating we only turned $n_z \rightarrow 0$, that is, $n_x$ and $n_y$ are considered up to high order in $1/\omega$ for this computation.  

Now, for ${\bf n}_{A^2}$ we have found the same lower integral in ``workMac10.txt" (with $M = N = 2$).

\begin{table}
\begin{tabular}{|c||c|c|} \hline
	integrand ($M$)	&	integral $Q_\var$ / $\Delta Q_\var$ &		integral $\Delta Q_\ex$ \\\hline
	$f_{\I}$		&	0.0991166116(5) [00]	& 3e-6	\\ \hline
	$f_{\I}$(1)	&	-6.77e-7 [1011]	&	\\
	$f_{\I}$(2)	&	-1.99e-6 [1013]	&	\\
	$f_{\I}$(3)	&	-2.00e-6 [1014/1015]	&	\\
	\hline
	$f_{\II}$		&	0.00633429(51) [000]	&	\\
	$f_{\II}$(1)	&	0.00631970(40) [204]	&	\\
	$f_{\II}$(2)	&	0.00632767(44) [206]	&	\\
	\hline
	$f_{\II}$		&	0.0063346 (13) [000]	&	 \\
	$f_{\II}$(1)	&	0.0063198(12) [204]	&	 \\
	$f_{\II}$(2)	&	0.0063279(12) [206]	&	 \\
	\hline
	$f_{\II}^2$		&	0.00111881851(1) [000]	&	\\
	$f_{\II}^2$(1)[*]	&	0.00111868926(1) [260]	&	\\
	$f_{\II}^2$(2)	&		[]&	\\
	\hline
\end{tabular}
\caption{Integral results for the full Gaussian pulse, which is taken in the interval $[-t_{\gate}, t_{\gate}]$.  For the computation of these numbers use ${\bf n}_{\var}$ $\mathcal{O}\left(A^3\right)$.  [*] Zero initial guess yielded no improvement.}
\label{integral_resultsFull}
\end{table}

\subsection{Integrals Over Full Pulse}
\label{integrand1_full}

Using the same gate duration of the half Gaussian pulse, $t_\text{gate} = 12\sigma$ with $\sigma = 2\times t_c = 2\times(2\pi)$, we now integrate from $t_0 = -t_\text{gate}$ to $t_\text{gate}$ over the full Gaussian pulse.  Accordingly, the values of the new integrals, which are shown in Table \ref{integral_resultsFull}, should be a bit more than twice as the old ones [see Table \ref{integral_results}].

\subsection{Sources of Errors}
\label{errorSources}

Here we list five possible sources of error:
\begin{enumerate}
	\item Computation of integral [this numerical uncertainty is given in parentheses, for example in Fig.~\ref{integral_results}]
	\item Errors due to the computation of the propagator [numerical values given in bottom section of Table \ref{integral_results}]
	\item Errors due to the numerical derivatives
	\item Approximations through simplistic choice of integrand
	\item ``Boundary conditions'' 
\end{enumerate}

(5)  The first four items listed here are under control (see Sec.~\ref{alreadyExplored} below).  The fifth item is made small by considering a pulse shape that goes to zero very smoothly.




\subsubsection{The Items Already Explored}
\label{alreadyExplored}

(1)  Done.

(2)  See discussion below in Sec.~\ref{errorAnalysis}.

(3)  We compute derivatives using a difference quotient.  For example, the first derivative of $\alpha$ is given by
\begin{eqnarray}
	\dot \alpha(t) \approx \frac{\alpha(t+h) - \alpha(t-h)}{2h}.
	\label{derivativeNumerical}
\end{eqnarray}
For the standard parameters [in particular, $A \sim 0.002$, $\sigma = 4\pi$, $t_c = 2\pi$], we have used $h = 10^{-3}$ and $h = 10^{-4}$ with the same results for the integral.  In case of the second derivative of $\alpha$ we have gotten (probably) bad results for the choice of $h = 10^{-5}$.  

(4)  We have compared the integral results for the respective ``simple'' integrands to those of the ``full'' integrands, and they do not pose a problem for the calculation [assuming the numerical derivatives are computed accordingly, e.g., $10^{-4} \leq h \leq 10^{-3}$ in Eq.~(\ref{derivativeNumerical})].  

\begin{eqnarray}
	\dot \alpha(t) &\approx& \frac{\alpha(t+h) - \alpha(t-h)}{2h}.
\end{eqnarray}

\subsection{Error Propagation}
\label{errorAnalysis}

Recall that for a function $f(a, b)$ with uncertainties $\Delta a$ and $\Delta b$ we find an uncertainty $\Delta f$ to be
\begin{eqnarray}
	\Delta f = \left |\frac{\partial f(a, b)}{\partial a} \right| \Delta a + \left |\frac{\partial f(a, b)}{\partial b} \right| \Delta b.
	\label{errorPropagation}
\end{eqnarray}

In our case, the function $f$ is either the integrand $f_{\I} = \lambda_+$ or $f_{\II} = \lVert \dot \H(t) \rVert$, which are given in Eqs.~(\ref{Qf_I-Final}) and (\ref{easyII}), respectively.  [Note that we have found a closed-form expression for $f_{\II}$ in Sec.~\ref{explicit_formulaII}.]  That is,
\begin{eqnarray}
	f_{\I}(\alpha, \dot \alpha, \dot {\hat n}) = \sqrt{\dot \alpha^2 + (1/2)|\dot {\hat n}|^2 [1 - \cos (2\alpha)]}.
	\label{fIIIErrorAnalysis}
\end{eqnarray}

Recall that we have another, simplified way of computing this integrand,
\begin{eqnarray}
	f_{\I}^{\text{simplified}} = | \dot {\bf n} |.
	\label{fIIISimErrorAnalysis}
\end{eqnarray}
I have compared the integrals $Q^{\text{simplified}} = \int f^{\text{simplified}}_{\I}$ in and $Q = \int f_{\I}$, and they both yield the numbers shown in Table \ref{integral_results} up to the given accuracy.  

In case of $f_{\II}$ we (currently) only use the simplified integrand, which is given by
\begin{eqnarray}
	f_{\II}^{\text{simplified}} = | \ddot {\bf n} |^2.
	\label{fIIErrorAnalysis}
\end{eqnarray}
It follows that we deal with an integrand ($f_{\I}$ or $f_{\I}^{\text{simplified}}$ or $f_{\II}^{\text{simplified}}$)
\begin{eqnarray}
	f_{\I}(\alpha, \dot \alpha, \dot {\hat n}) \pm \Delta f_{\I}.
\end{eqnarray}
Let us now determine $\Delta f_{\I}$.

\subsubsection{What We Know}

I think I can assume I know the error in the the vector ${\bf n}$, denoted $\Delta n$, i.e., the true value of $n(t, \beta_0) = |{\bf n}(t, \beta_0)|$ lies somewhere within the interval
\begin{eqnarray}
	[n(t, \beta_0) - \Delta n, n(t, \beta_0) + \Delta n].
	\label{nDelta}
\end{eqnarray}
Given $\alpha = |{\bf n}|$ and $\hat n = {\bf n}/\alpha$, we further define error quantities
\begin{eqnarray}
	\Delta \alpha = \alpha - \alpha_0, \text{ and } \Delta \hat n = \hat n - \hat n_0,
	\label{DANH}
\end{eqnarray}
where we take $\alpha$ and $\hat n$ to correspond to the effective time evolution, while $\alpha_0$ and $\hat n_0$ 

To obtain a numerical estimate of the quantitites (\ref{DANH}), let us define the time evolution operators for the effective and the exact trajectories:
\begin{eqnarray}
	U_{\eff}(\alpha, \hat n) &=& e^{-i\alpha \hat n \cdot \sigma} = I \cos(\alpha) -i \hat n\cdot \sigma \cos(\alpha),
	\label{UEff} \\
	U_{\ex}(\alpha, \hat n) &=& \mathcal T e^{-i\int d\tau H_{\ex}(\tau)} \equiv e^{-i\alpha_{\ex} \hat n_{\ex} \cdot \sigma}.
\end{eqnarray}
Note that we left the dependence on time $t$ and the gauge parameter $\beta_{\ex}$ implicit, e.g., $\alpha = \alpha(t, \beta_{\ex})$.  We also define the difference between these two operators,
\begin{eqnarray}
	\Delta U = U_{\eff} - U_{\ex}.
\end{eqnarray}

Note that from the difference $\Delta U$ we could determine the error quantity $\Delta$ [cf.~(\ref{DANH})] by writing
\begin{eqnarray}
	\Delta_c &\equiv& \frac12 \text{tr} \Delta U = \cos\alpha - \cos\alpha_{\ex}  = \cos(\alpha_{\ex} + \Delta \alpha) - \cos\alpha_{\ex},
\end{eqnarray}
and the somehow solving for $\Delta \alpha$.  

\subsubsection{Simple Way}

Recall, however, that we already have functions for the ``effective'' quantities $\alpha$ and $\hat n$ [search for `def alpha' and 'def nHat' in fromMathematica.py], so in order to find the desired errors we merely need to write functions for $\alpha_{\ex}$ and $\hat n_{\ex}$ based on $U_{\ex}$.  Note that the function for $U_{\ex}$ can be obtained from the exact time evolution psi(t, beta0, ...) [search for `def psi' in fromMathematica.py] by choosing the input parameter \emph{out} = 'thetaPhi.'  

As described above, we can also easily obtain the equivalent exact quantities.  For this, note that Eq.~(\ref{UEff}) implies that
\begin{eqnarray}
	\cos(\alpha_{\ex}) &=& \frac12 \text{tr} \ U_{\ex}, \\
	\Rightarrow -i n_j \sin(\alpha_{\ex}) &=& \frac12 \text{tr} \ U_{\ex} \sigma_j \qquad \ \qquad (j = x, y, z)  \\
	\Rightarrow \qquad \qquad n_j &=& \frac{i}{2\sin(\alpha_{\ex})} \text{tr} U_{\ex} \sigma_j.
\end{eqnarray}
These quantities are computed in fromMathematica.py (search for `def alphaNHatExact').

\subsubsection{Now compute the error}

I believe we can figure out an estimate of $\Delta n_x$, $\Delta n_y$ and $\Delta n_z$.  Assuming this is given, we can compute the error in $\alpha$, which is length of the $n$-vector,
\begin{eqnarray}
	\alpha &=& |{\bf n}| = \sqrt{n_x^2 + n_y^2 + n_z^2}, \\
	\Rightarrow \frac{\partial \alpha}{n_i} &=& \frac{n_i}\alpha \qquad (\text{for } i = x, y, z), 
	\label{alphaDerivative}\\
	\stackrel{(\ref{errorPropagation})}{\Rightarrow}
	 \Delta \alpha &=& \sum_{i} \frac{n_i}{\alpha} \Delta n_i.
	\label{DeltaAlpha}
\end{eqnarray}

Recall that we also have
\begin{eqnarray}
	\hat n &=& \frac{\bf n}{n} = \frac{\bf n}{\alpha}.\\
	\frac{\partial }{\partial \alpha} \frac 1{\alpha} &=& -\frac 1{\alpha^2} \\
	\stackrel{(\ref{errorPropagation})}{\Rightarrow}
	 \Delta \hat n & = & \frac{\Delta \bf n}{\alpha} + \frac{\Delta \alpha}{\alpha^2} {\bf n},
\end{eqnarray}
where $\Delta \alpha$ is given above in Eq.~(\ref{DeltaAlpha}).

\subsubsection{Time Derivatives}

It is, of course, possible that some error, say $\Delta \alpha$ or $\Delta {\bf n}$, probably builds up (he didn't use this particular verb) over a Magnus interval $t_c$, so that we have
\begin{eqnarray}
  \Delta \dot {\bf n} = \frac{\Delta {\bf n}}{t_c}, \\
  \Delta \dot \alpha = \frac{\Delta \alpha}{t_c},		\label{DeltaA}\\
  \Delta \dot {\hat n} = \frac{\Delta {\hat n}}{t_c}.	\label{DeltaNH}
\end{eqnarray}
From Eq.~(\ref{fIIIErrorAnalysis}) we take that the latter two equations given here are required, i.e., (\ref{DeltaA}) and (\ref{DeltaNH}), as well as the uncertainty in $\alpha$ itself, i.e., Eq.~(\ref{DeltaAlpha}).

\subsubsection{First Results}

Using the function ``deltaAlphaNHat'' (see fromMathematica.py) I computed the errors of the $n$-vector for various approximations of the propagator [for this I used dbl\_test.py].  The results are noted in Table \ref{integral_results}, lowest rows.  For the most accurate treatment, in which we started with the effective Hamiltonian $\H_{\eff}$ of order $1/\omega^5$, and where we kept terms $\propto A^3$ in the Taylor expansion, we found $|\Delta \hat n| \lesssim 10^{-8}$, $\Delta \alpha \lesssim 10^{-10}$, perhaps most importantly,
\begin{eqnarray}
	\Delta |{\bf n}| \equiv \Delta n \lesssim 1.4 \times 10^{-9}.
\end{eqnarray}
These are absolute (not relative) numbers.  

Using the three quantities $\Delta \alpha$, $\Delta \dot \alpha$ and $\Delta \hat n$ in hand we can thus compute the error for $f_{\I}$ given in Eq.~(\ref{fIIIErrorAnalysis})
\begin{eqnarray}
	\Delta f_{\I} &=& \sum_{p = \alpha, \dot \alpha, |\dot {\hat n}|} \frac{\partial f_{\I}}{\partial p} \times \Delta p \\
			&=& \frac{1}{2 f_{\I}} \left[
			2\dot \alpha \times \Delta \dot \alpha + (1/2)[1 - \cos (2\alpha)] (2|\dot {\hat n}|) \times \Delta |\dot {\hat n}| + (1/2)|\dot {\hat n}|^2 |2\sin(2\alpha)| \times \Delta \alpha
			\right]  \nonumber \\
			&=& \frac{1}{2 f_{\I}} \left[
			2\dot \alpha \times \Delta \dot \alpha + [1 - \cos (2\alpha)] |\dot {\hat n}| \times \Delta |\dot {\hat n}| + |\dot {\hat n}|^2 \ |\sin(2\alpha)| \times \Delta \alpha
						\right].
\end{eqnarray}
The integral over this quantity [computed using ``dbl\_plotting.py''] yields the following uncertainty of the integral $Q$,
\begin{eqnarray}
	\Delta Q_{\I} = 7.2(1)\times 10^{-6}\text{(*)}.
\end{eqnarray}
(*)  This number is not up to date [July 11th, 2019].  Use the simplified result below in Eq.~(\ref{QIIISim}) for now.  

In the case of the ``simplified'' integrand $f_{\I}^{\text{simplified}}$ given in Eq.~(\ref{fIIISimErrorAnalysis}) we need to use $\Delta \dot n \approx \Delta n/t_c$ (with the absolute uncertainty $\Delta n \lesssim 1.4\times 10^{-9}$ (see Table \ref{integral_results}), so we can compute
\begin{eqnarray}
	\Delta f_{\I}^{\text{simplified}} &=& \frac{\partial f^{\text{simplified}}_{\I}}{\partial |\dot n|} \times \Delta |\dot {\bf n}| \nonumber = \Delta \dot n.
\end{eqnarray}
Similar to above, the integral over this quantity (computed using ``dbl\_plotting.py'') yields an uncertainty of the integral $Q^{\text{simplified}}$ that is given by
\begin{eqnarray}
	\Delta Q_{\I}^{\text{simplified}} = 1.06 \times 10^{-7} \pm \mathcal O(10^{-9}).
	\label{QIIISim}
\end{eqnarray}

In the case of the ``simplified'' integrand $f_{\II}^{\text{simplified}}$ given in Eq.~(\ref{fIIErrorAnalysis}) we will have to use $\Delta \ddot n = \Delta n/t_c^2$ (I suppose) in order to compute
\begin{eqnarray}
	\Delta f_{\II}^{\text{simplified}} &=& \frac{\partial f^{\text{simplified}}_{\II}}{\partial |\ddot n|} \times \Delta |\ddot {\bf n}| = \Delta \ddot n.
\end{eqnarray}
From Eq.~(\ref{QIIISim}) we find directly that since $\Delta \ddot n = \Delta n/t_c^2 = \Delta \dot n/t_c$, we have
\begin{eqnarray}
	\Delta Q_{\II}^{\text{simplified}} = (1.1/2\pi) \times 10^{-7} \pm \mathcal O(10^{-9}) = 1.75 \times 10^{-8}.
	\label{QIISim}
\end{eqnarray}

The bigger problem is that I believe I cannot use the simplified integrand for $f_{\II}$.  

\subsubsection{The Case Against Integrand I}
\label{caseAgainstI}

From Table \ref{integral_results} we learn the following.  The variational approach yields a set of variational parameters that result in a smaller integral $Q$.  The corresponding trajectory, when plotting it for a given value of $\beta_0$, basically takes a slight shortcut compared to ``our'' original effective trajectory.  

\begin{table}
\begin{tabular}{|c|c|}\hline
	integrand & integral result $Q_i$ with $i = 0, \var$ \\\hline\hline 
	digits	&	0.123456789(00) [00] \\
	$\lambda_{+, 0}$	&	0.039826282(13) [00] \\
	$\lambda_{+, \var}$	&	0.039825489(30) [10] \\\hline
	digits	&	0.12345678901(00) [000] \\
	$f_{\II, 0}$		&		0.00304297287(50) [000]\\
	$f_{\II, \var}$	&		0.00304289075(60) [212]\\\hline
\end{tabular}
\caption{Most informative integral results taken from Table \ref{integral_results}.}
\label{integral_results_informative}
\end{table}



We have more-or-less certainly confirmed that the integral result $Q_{\var}$ is indeed smaller than $Q_0$, the integral for the original Hamiltonian.  Consulting Table \ref{integral_results_informative}, the difference
\begin{eqnarray}
	\Delta Q = Q_0 - Q_{\var} = 7.8\times 10^{-7}.
	\label{caseIII}
\end{eqnarray}
Here we have taken into account the inaccuracies due to the numerical integration.  Now, the uncertainty computed in Eq.~(\ref{QIIISim}) is only $1.06 \times 10^{-7}$, which strongly suggests that the $\Delta Q$ found by us is significant.

\section{Conclusions}
\label{conclusions}

The main statement of Axiom \ref{axiom3_axiom} is that the effective Hamiltonian results in the lowest integral value of the functional $Q = \int_{\beta_0} \int_{\tau} f \d \tau \d \beta_0$.  Here, an integrand $f(t, \beta_0)$ dependent on time $t$ and the gauge parameter $\beta_0$, is integrated over both of its values for a full single-qubit drive pulse.  In these notes we have analytically and numerically analyzed this axiom for the integrand $f = f_\I = \eig_+(\H)$, or the positive eigenvalue of the Hamiltonian, introduced in Sec.~\ref{integrands}.  

In Sec.~\ref{caseAgainstI} we have noted that the integral $Q$ for the effective Hamiltonian is significantly larger than that of the ``best'' variational Hamiltonian.  This implies that our proposed axiom is violated for the integrand $f_\I$.  

Given the prominent qualitative differences between the effective and exact qubit trajectories (cf.~Fig.~\ref{trajectories}), we expect that some other definition for the effective Hamiltonian of the exact rotating wave approximation---beyond infinite series with unclear convergence behavior---may be found.

\section{Acknowledgements}

We thank Evangelos Varvelis, Alwin van Steensel, Veit Langrock, Fabian Hassler and Cica Gustiani for many useful discussions.

\newpage

\appendix

\section{Magnus Expansion}
\label{magnus_appendix}

The Magnus expansion \cite{magnus1954, ernst87, waugh07} is a method used for time-dependent perturbation theory.  The basic idea is to write the time evolution operator, which generally requires the time ordering operator, as a true exponential function of an operator that is to be determined perturbatively.  

Consider the stroboscopic time evolution for the set of times $\{t_0, t_0 \pm t_c, \ldots \}$, as given in Eq.~(\ref{sets}), with the drive period $t_c=\pi/\omega$ and the time offset $t_0\in[0, t_c)$.  The stroboscopic time evolution operator parallel to that in Eq.~(\ref{goal0}) can then be written as
\begin{equation}
	U_{t_0}(t_0 + n t_c, t_0) = e^{-i \overline{\mathcal H} n t_c},
	\label{UM2}
\end{equation}
for integers $n$.  The Magnus expansion $\overline \H$ can be written as a series,
\begin{equation}
	\overline \H = \sum_{k=0}^\infty \overline \H^{(k)}.
	\label{MagnusSeries}
\end{equation}
The three lowest-order terms $\H^{(k)}$ with $k = 0$, 1 and 2, read
\begin{eqnarray}
	\overline \H^{(0)} &=& \frac1{n t_c} \int_{t_0}^{t_0 + nt_c} \text d \tau \H(\tau), \label{Hb0}\\
	\overline \H^{(1)} &=& \frac{-i}{2 nt_c} \int_{t_0}^{t_0 + nt_c}\text d \tau' \int_{t_0}^{\tau'} \text d \tau [\H(\tau'),\H(\tau)], \label{Hb1}\\
	\overline \H^{(2)} &=& -\frac{1}{6 nt_c} \int_{t_0}^{t_0 + nt_c}\text d \tau'' \int_{t_0}^{\tau''} \text d \tau' \int_{t_0}^{\tau'} \text d \tau \Big\{[\H(\tau''),[\H(\tau'),\H(\tau)]] + [[\H(\tau''),\H(\tau')],\H(\tau)]\Big\}. \quad \label{Hb2}
\end{eqnarray}

Terms of higher order may be determined recursively, see, e.g., Refs.~\cite{blanes09,blanes10}.


\section{Various supporting calculations}
\label{support}


\subsection{Derivatives of trial functions}
\label{explicit_derivatives}

To write the first and second derivatives of the vector $n_\var$ given in Eq.~(\ref{trial_fxns}), we defined the outer factor $f_{\text{outer}} = [\sin(\theta_t)^2-\sin(\eta\phi)^2] e^{a t^2}$ where $\theta_t = \omega (t-\beta_0)+\eta\phi$ (we argued above that $\Omega = \omega = 1/2$) and $a=-(1+c)/(2 \sigma^2)$.  The derivatives of this outer factor are then given by
\begin{eqnarray}
	\dot f_{\text{outer}} &=& 2\omega\sin(\theta_t)\cos(\theta_t)e^{a t^2} + 2at \ f_{\text{outer}} \nonumber \\
			&=& \omega\sin(2\theta_t)e^{a t^2} + (2at)f_{\text{outer}},
	\label{outerPrime} \\
	\ddot f_{\text{outer}} &=&  2[\omega^2(\underbrace{\cos \theta_t^2 - \sin\theta_t^2}_{\cos2\theta_t}) + 2a\omega t\underbrace{\sin\theta_t \cos\theta_t}_{\tfrac12 \sin2\theta_t}]e^{a t^2}\nonumber \\
			&& + 2a f_{\text{outer}} + 2at \dot f_{\text{outer}} \nonumber \\
			&=&  2[\omega^2\cos2\theta_t + a\omega t \sin2\theta_t] e^{a t^2} + 2a f_{\text{outer}} + 2at \dot f_{\text{outer}}.
	\label{outerPrimePrime}
\end{eqnarray}
To avoid mistakes as much as possible, these equations were checked analytically using Mathematica, and their Python code has been checked numerically.  Numerical checks can be done using \texttt{wrongFactor} and \texttt{wrongFactor2} in \texttt{trial}, and plotting $f_{\I}$ or $f_{\II}$ in plotting.py (with \texttt{derivative1FDQ} =  \texttt{True}).


\subsection{Time derivative of operator exponential}
\label{Udot}

As discussed in the beginning of Sec.~\ref{minimizeEigenvalue}, the derivative of a function $\exp(f(t))$ for an operator $f(t)$ is not trivial if $[f(t), \dot f(t)] \neq 0$.  Evangelos Varvelis pointed me to Ref.~\cite{blanes09}, in which Eqs.~(33)-(35) contain various forms of the derivative of the exponential of the time evolution operator.  Here, we choose Eq.~(35) from Ref.~\cite{blanes09},
\begin{eqnarray}
	\partial_t U = \partial_t e^{\Omega(t)} = \int e^{s \Omega(t)} (\partial_t \Omega(t)) e^{(1-s)\Omega(t)}\d s
	\label{derivative}
\end{eqnarray}
with the Magnus expansion $\Omega(t)$.  

Using
\begin{eqnarray}
	\bar \H(t) = i \Omega(t) = {\bf n}(t)\cdot \sigma = \alpha(t) \hat n(t)\cdot \sigma,
	\label{Hbar}
\end{eqnarray}
the computation based on Eq.~(\ref{derivative}) reads
\begin{eqnarray}
	Q	&\stackrel{(*)}{=}& \int \int \eig _+\left[
		\int_0^1 
		e^{i s \bar \H(t)} (\partial_t \bar \H)
		e^{i(1-s) \bar \H(t)} e^{- i\bar \H(t)} \d s
		\right] \d t \d \beta_0
		 \nonumber\\
		&=& \int \int \eig _+\left[
		\int_0^1 
		e^{i s \bar \H(t)} (\partial_t \bar \H)
		e^{-is \bar \H(t)} \d s
		\right] \d t \d \beta_0.
\end{eqnarray}
Substituting for $\bar \H$ as given in Eq.~(\ref{Hbar}),
\begin{eqnarray}
		Q &=& \int \int \eig _+\left[
			\int_0^1 
		e^{is \bar \H(t)}(\partial_t \alpha \hat n \cdot \sigma )
		e^{-is \bar \H(t)} \d s
		\right] \d t \d \beta_0 \nonumber\\
		&=& \int \int \eig _+\left[ \dot \alpha \hat n\cdot \sigma + 
		\alpha \underbrace{\int_0^1  e^{is \bar \H(t)}(\dot {\hat n} \cdot \sigma) e^{-is \bar \H(t)}
		\d s}_{\equiv I} \right] \d t \d \beta_0.
	\label{alphaTimesNHat} 
	\label{minimize}
\end{eqnarray}

\subsection{Algebra of Unit Vectors}
\label{unitVectors}

\subsubsection{No Approximation}
\label{unit_vectors_elegant}

To properly compute the second derivative of a unit vector, I believe I we can do the following.  First define the velocity vector $\hat n_v$, which points along the direction of the velocity,
\begin{eqnarray}
	\vec n_v  &=& \partial_t {\hat n}.
	\label{def_nv}
\end{eqnarray}
Using $\vec n_v \equiv n_v \hat n_v$ and $\partial_t \hat n = |\dot {\hat n}| \hat n_v$, it is clear that
\begin{eqnarray}
	\partial_t {\hat n} &\equiv& n_v \hat n_v, \qquad \qquad n_v = |\dot {\hat n}|.
	\label{nHatD}
\end{eqnarray}
Then take another derivative,
\begin{eqnarray}
	\partial_t^2 \hat n &\stackrel{(\ref{def_nv})}{=}& \partial_t \vec n_v \\
			&\stackrel{(\ref{nHatD})}{=}& \dot n_v \hat n_v + n_v \dot {\hat n}_v. 
	\label{nHatDD}
\end{eqnarray}

The computation of $\dot {\hat n}_v$ looks as follows [we have tried to gain some insights using the (second part within the) Mathematica notebook 'integrand4.nb' in the github repository [cf.~Sec.~\ref{coding}] in the folder \texttt{/exactRWA/programs/variational\_minimization}].  First define a slightly unintuitive acceleration vector, which is the temporal derivative of the \emph{unit vector} ${\hat n}_v$, i.e.,
\begin{eqnarray}
	\vec n_a = \dot {\hat n}_v = a \hat n + b \hat n_{\perp}
	\label{nVHatDot}
\end{eqnarray}
where $\vec n_a = n_a \hat n_a$.  We define $\hat n_{\perp} = \hat n \times \hat n_v$, which guarantees $\hat n_{\perp} \perp \hat n$ and $\hat n_{\perp} \perp \hat n_v$.  We wish to express the integrand $f_\II$ in terms of the three unit vectors
\begin{eqnarray}
	\left\{\hat n, \hat n_v, \hat n_{\perp} \right \},
	\label{system}
\end{eqnarray}
which are mutually perpendicular to one another.  We then find
\begin{eqnarray}
	a \stackrel{(\ref{nVHatDot})}{=} \vec n_a \cdot \hat n,
	\label{littlea}
\end{eqnarray}
and together with $n_a = |\vec n_a| \stackrel{(\ref{nVHatDot})}{=} a^2 + b^2$ we have
\begin{eqnarray}
	b = \pm \sqrt{n_a^2 - a^2} = \pm \sqrt{n_a^2 - \vec n_a \cdot \hat n}.
	\label{littleb_Bad}
\end{eqnarray}
However, now we don't know the sign of $b$ so it's probably better to compute $b$ the same way as we computed $a$ in Eq.~(\ref{littlea}),
\begin{eqnarray}
	b \stackrel{(\ref{nVHatDot})}{=} \vec n_a \cdot \hat n_\perp.
	\label{littleb}
\end{eqnarray}

To be clear, the norm of the vector $\vec n_a$ is $n_a \stackrel{(\ref{nVHatDot})}{=} |\dot {\hat n}_v|$, and combining Eqs.~(\ref{nVHatDot}), (\ref{littlea}) and (\ref{littleb}) we obtain
\begin{eqnarray}
	\vec n_a = (\vec n_a \cdot \hat n)\  \hat n +  (\vec n_a \cdot \hat n_\perp) \ \hat n_{\perp}.
	\label{vec_a}
\end{eqnarray}

We can thus write the second derivative of the vector $\hat n$ as follows,
\begin{eqnarray}
	\partial_t^2 \hat n &\stackrel{(\ref{nHatDD}), (\ref{nVHatDot})}{=}& \dot n_v \hat n_v + n_v \vec n_a \label{bad} \\
			&\stackrel{(\ref{vec_a})}{=}& \dot n_v \hat n_v + n_v \left[(\vec n_a \cdot \hat n) \ \hat n + (\vec n_a \cdot \hat n_\perp) \ \hat n_{\perp} \right].
	\label{ddotnHatApp}
\end{eqnarray}
Note that the reason why Eq.~(\ref{ddotnHat}) should be more desirable than Eq.~(\ref{bad}) is that it is expressed using the unit vectors given in Eq.~(\ref{system}).

%
%


\subsection{Eigenvalue of (Traceless) 2 by 2 Matrix}

The eigenvalues $\pm \lambda$ of a traceless $2\times 2$ matrix $M_2$ can be computed quite easily using the determinant.  This becomes clear (recall that the determinant is basis-independent) as follows,
\begin{eqnarray}
	\det M_2 = \det \text{diag} (\lambda, -\lambda) = -\lambda^2,
	\label{det}
\end{eqnarray}
because of which we have
\begin{eqnarray}
	\eig_+ M_2 = \sqrt{\lambda^2} \stackrel{(\ref{det})}{=} \sqrt{-\det M_2}.
	\label{eigenvalueDet}
\end{eqnarray}


\section{Explicit formulas for effective Hamiltonian}
\label{effH_appendix}

In this appendix we give the explicit formulas for Hamiltonian series (\ref{HeffSeries}) up to order $1/\omega^3$,
\begin{eqnarray}
	\H_{\eff}(t; \beta_0) &=& \sum_{k=0}^{\infty} h_{k}(t; \beta_0) (1/\omega)^k \nonumber \\ &=& \H_0(t; \beta_0) + \H_1(t; \beta_0) + \H_2(t; \beta_0) + \H_3(t; \beta_0) + \H_4(t; \beta_0) \nonumber \\
	&& + \H_5(t; \beta_0) + \O(1/\omega^4)
	\label{HEffective}
\end{eqnarray}
with $\H_i = h_i/\omega^i$ for $i = 0, 1, 2, 3$.  For readability, below all dependencies on time and the gauge parameter $\beta_0$ of the Hamiltonian or the envelope functions are kept implicit.  These formulas have been obtained assuming the on-resonant rotating-frame Hamiltonian (\ref{Hrot}), and have been determined following the recurrence procedure of Ref.~\cite{zeuch18}.  

The lowest-order Hamiltonian is simply given by the Hamiltonian of the standard rotating wave approximation, $\H_0 = \H_\RWA = (H_1/4)\sigma_x$, as also given in Eq.~(\ref{HRWA}).

\subsection{Time-dependent drive envelope}
\label{HEff_genericH1}

For a generic envelope $H_1(t)$, the lowest three corrections are
\begin{eqnarray}
	\mathcal H_1 &=& \frac{H_1^2}{32 \omega}(1-2 \cos\beta_0)\sigma_z+ \frac{\dot H_1}{8 \omega} [\sin\beta_0\sigma_x+\cos\beta_0\sigma_y], \label{backslashH1}\\
	\mathcal H_2 &=& \frac{H_1^3}{256 \omega ^2}[(-2 + 2 \cos\beta_0-\cos(2 \beta_0))\sigma_x+(2 \sin\beta_0+\sin(2 \beta_0))\sigma_y] \nonumber \\
				&& + \frac{3 H_1 \dot H_1}{32 \omega ^2} \sin\beta_0\sigma_z+ \frac{\ddot H_1}{16 \omega ^2}[\cos\beta_0\sigma_x-\sin\beta_0\sigma_y],
\end{eqnarray}
together with
\begin{eqnarray}
	\H_3		&=& \frac{H_1^4}{2048 \omega ^3}(1-2 \cos(\beta_0)-3 \cos(2 \beta_0))\sigma_z \nonumber \\
				&& + \frac{H_1^2 \dot H_1}{1024 \omega ^3}[(9 \sin(2 \beta_0)-12 \sin(\beta_0))\sigma_x+(36 \cos(\beta_0)+9 \cos(2 \beta_0)-8)\sigma_y] \nonumber\\
				&& +\frac{\dot H_1^2}{128 \omega ^3}(6 \cos(\beta_0)+1)\sigma_z + \frac{H_1 \ddot H_1}{64 \omega ^3}(4 \cos(\beta_0)-1)\sigma_z \nonumber \\
				&& - \frac{\dddot H_1}{32 \omega ^3}[\sin(\beta_0)\sigma_x+\cos(\beta_0)\sigma_y].
\end{eqnarray}
The corrections of order $1/\omega^4$,
\begin{eqnarray}
	\H_4 &=& \frac{H_1^5}{16384 \omega ^4}[(5 \cos(\beta_0)-\cos(2 \beta_0)-\cos(3 \beta_0)-9)\sigma_x+(5 \sin(\beta_0)+4 \sin(2 \beta_0)+\sin(3 \beta_0))\sigma_y] \nonumber \\ 
	&& + \frac{45 H_1^3 \dot H_1}{8192 \omega ^4}(2 \sin(\beta_0)+\sin(2 \beta_0))\sigma_z \nonumber \\ 
	&& + \frac{5 H_1 \dot H_1^2}{2048 \omega ^4}[(3 \cos(2 \beta_0)-4 \cos(\beta_0))\sigma_x - (20 \sin(\beta_0)+3 \sin(2 \beta_0))\sigma_y] \nonumber \\ 
	&&  +\frac{5 H_1^2 \ddot H_1}{4096 \omega ^4}[(-8 \cos(\beta_0)+5 \cos(2 \beta_0)+8)\sigma_x-(24 \sin(\beta_0)+5 \sin(2 \beta_0))\sigma_y]  \nonumber \\ 
	&& -\frac{5 \dot H_1 \ddot H_1 \sin(\beta_0) \sigma_z}{64 \omega ^4} - \frac{5 H_1 \dddot H_1 \sin(\beta_0)}{128 \omega ^4}\sigma_z+ \frac{H_1^{(4)}}{64 \omega ^4}[-\cos(\beta_0)\sigma_x+\sin(\beta_0)\sigma_y],
\end{eqnarray}
and of order $1/\omega^5$,
\begin{eqnarray}
	\H_5 &=& \frac{H_1^6}{786432 \omega ^5}(18 \cos(\beta_0)-60 \cos(2 \beta_0)-10 \cos(3 \beta_0)-9)\sigma_z \nonumber \\ 
	&& + \frac{H_1^4 \dot H_1}{196608 \omega ^5}[(-285 \sin(\beta_0)+150 \sin(2 \beta_0)+55 \sin(3 \beta_0))\sigma_x \nonumber \\ 
		&& +(825 \cos(\beta_0)+330 \cos(2 \beta_0)+55 \cos(3 \beta_0)-297)\sigma_y] \nonumber \\ 
	&& + \frac{H_1^2 \dot H_1^2}{32768 \omega ^5}(1000 \cos(\beta_0)+285 \cos(2 \beta_0)-104)\sigma_z\nonumber \\ 
	&& + \frac{\dot H_1^3}{8192 \omega ^5}[(40 \sin(\beta_0)-15 \sin(2 \beta_0))\sigma_x+(-200 \cos(\beta_0)-15 \cos(2 \beta_0)-24)\sigma_y] \nonumber \\ 
	&& +\frac{3 H_1^3 \ddot H_1}{16384 \omega^5}(65 \cos(\beta_0)+25 \cos(2 \beta_0)-16)\sigma_z\nonumber \\ 
	&& + \frac{H_1 \dot H_1 \ddot H_1}{8192 \omega ^5}[(160 \sin(\beta_0)-95 \sin(2 \beta_0))\sigma_x+(-800 \cos(\beta_0)-95 \cos(2 \beta_0)+72)\sigma_y]\nonumber \\ 
	&& + \frac{\ddot H_1^2}{512 \omega ^5}(1-20 \cos(\beta_0))\sigma_z+ \frac{H_1^2 \dddot H_1}{16384 \omega ^5}[(80 \sin(\beta_0)-65 \sin(2 \beta_0))\sigma_x+\nonumber \\ 
	&& (-400 \cos(\beta_0)-65 \cos(2 \beta_0)+64)\sigma_y]+ \frac{\dot H_1 \dddot H_1}{256 \omega ^5}(-15 \cos(\beta_0)-1)\sigma_z \nonumber \\ 
	&& + \frac{H_1 H_1^{(4)}}{256 \omega ^5}(1-6 \cos(\beta_0))\sigma_z+ \frac{H_1^{(5)}}{128 \omega ^5}[\sin(\beta_0)\sigma_x+\cos(\beta_0)\sigma_y],
\end{eqnarray}
are also used in our calculation.

\subsection{Constant drive envelope}
\label{HEff_constantH1}

For a constant amplitude $H_1(t) = H_1$ the formulas above simplify as follows,
\begin{eqnarray}
	\H_1 &=& \frac{H_1^2}{32 \omega}(1-2 \cos(\beta_0))\sigma_z, \\
	\H_2 &=& \frac{H_1^3}{256 \omega ^2}[(2 \cos(\beta_0)-\cos(2 \
	\beta_0)-2)\sigma_x+(2 \sin(\beta_0)+\sin(2 \beta_0))\sigma_y], \\
	\H_3 &=& \frac{H_1^4}{2048 \omega ^3}(-2 \cos(\beta_0)-3 \cos(2 \
	\beta_0)+1)\sigma_z, \\
	\H_4 &=& \frac{H_1^5}{16384 \omega ^4}[(5 \cos(\beta_0)-\cos(2 \beta_0)-\cos(3 \
	\beta_0)-9)\sigma_x + \nonumber \\
		&& \qquad \qquad \qquad \qquad \qquad (5 \sin(\beta_0)+4 \sin(2 \beta_0)+\sin(3 \
	\beta_0))\sigma_y], \\
	\H_5 &=& \frac{H_1^6}{786432 \omega ^5}(18 \cos(\beta_0)-60 \cos(2 \beta_0)-10 \
	\cos(3 \beta_0)-9)\sigma_z.
\end{eqnarray}

We furthermore give the Hamiltonian coefficients $\H_6$ and $\H_7$ for constant driving when considering the next two orders [not explicitly shown in Eq.~(\ref{HEffective})],
\begin{eqnarray}
	\H_6 &=& \frac{H_1^7}{37748736 \omega ^6}[(252 \cos(\beta_0)+84 \cos(2\beta_0)-120 \cos(3 \beta_0)-15 \cos(4 \beta_0)-1224)\sigma_x \nonumber \\
		&& \quad \qquad \qquad \qquad \qquad + (252 \sin(\beta_0)+336 \sin(2 \beta_0)+160 \sin(3 \beta_0)+15 \sin(4 \beta_0))\sigma_y], \\
	\H_7 &=& \frac{H_1^8}{1811939328 \omega ^7}(10152 \cos(\beta_0)-4368 \cos(2 \beta_0)-1540 \cos(3 \beta_0)-105 \cos(4 \beta_0)-5076)\sigma_z. \nonumber\\
\end{eqnarray}

\bibliography{bibliography_variational}

\end{document}